\documentclass[11pt]{article}
\pdfoutput=1 % if your are submitting a pdflatex (\textit{i.e.} if you have
\usepackage{jheppub}
\usepackage[utf8]{inputenc}

\usepackage{graphicx}
\usepackage{caption,epigraph}
\usepackage{subcaption}
\usepackage{hyperref}
\usepackage[T1]{fontenc}
\usepackage[export]{adjustbox}
\usepackage[table]{xcolor}

\usepackage{amssymb}% http://ctan.org/pkg/amssymb
\usepackage{pifont}% http://ctan.org/pkg/pifont

\newcommand{\be}{\begin{equation}}
\newcommand{\ee}{\end{equation}}
\newcommand{\bea}{\begin{eqnarray}}
\newcommand{\eea}{\end{eqnarray}}

\usepackage{lipsum}

\newcommand\blfootnote[1]{%
  \begingroup
  \renewcommand\thefootnote{}\footnote{#1}%
  \addtocounter{footnote}{-1}%
  \endgroup
}

\DeclareUnicodeCharacter{2212}{-}
\usepackage{multirow, graphicx,amssymb,url,mathrsfs,amsmath}
\usepackage{amsxtra,amstext,latexsym,dsfont,amsfonts}
\usepackage{color,eucal}
\usepackage{float}
\usepackage{colortbl}
\usepackage{multirow}
\usepackage{tikz}
\usetikzlibrary{tikzmark}
\usetikzlibrary{arrows}
\usepackage{tabularx, array}
\newcolumntype{L}[1]{>{\raggedright\arraybackslash}p{#1}}
\newcolumntype{C}[1]{>{\centering\arraybackslash}p{#1}}
\newcolumntype{R}[1]{>{\raggedleft\arraybackslash}p{#1}}
\makeatother
\makeatletter
\newcommand{\myBig}{\bBigg@{1.75}}
\makeatother

\begin{document} 
%%%
\title{\LARGE Chiral magnetic waves in strongly coupled Weyl semimetals} 

\author[a,b,c]{Yongjun Ahn,}
\author[a,b,c]{Matteo Baggioli,}
\author[d,e]{Yan Liu,}
\author[a,b,c,*]{Xin-Meng Wu}\blfootnote{*Corresponding author.}

\affiliation[a]{School of Physics and Astronomy, Shanghai Jiao Tong University, Shanghai 200240, China}
\affiliation[b]{Wilczek Quantum Center, School of Physics and Astronomy, Shanghai Jiao Tong University, Shanghai 200240, China}
\affiliation[c]{Shanghai Research Center for Quantum Sciences, Shanghai 201315, China}
\affiliation[d]{Center for Gravitational Physics, Department of Space Science,
Beihang University, Beijing 100191, China}
\affiliation[e]{Peng Huanwu Collaborative Center for Research and Education, Beihang University, Beijing 100191, China}

\emailAdd{yongjunahn@sjtu.edu.cn}
\emailAdd{b.matteo@sjtu.edu.cn}
\emailAdd{yanliu@buaa.edu.cn}
\emailAdd{xinmeng.wu@sjtu.edu.cn}

\abstract{Propagating chiral magnetic waves (CMW) are expected to exist in chiral plasmas due to the interplay between the chiral magnetic and chiral separation effects induced by the presence of a chiral anomaly. Unfortunately, it was pointed out that, because of the effects of electric conductivity and dissipation, CMW are overdamped and therefore their signatures are unlikely to be seen in heavy-ion collision experiments and in the quark gluon plasma. Nonetheless, the chiral anomaly plays a fundamental role in Weyl semimetals and their anomalous transport properties as well. Hence, CMW could be potentially observed in topological semimetals using table-top experiments. By using a holographic model for strongly coupled Weyl semimetals, we investigate in detail the nature of CMW in presence of Coulomb interactions and axial charge relaxation and estimate whether, and in which regimes, CMW could be observed as underdamped collective excitations in topological materials.
}

\maketitle
%%%%%%%%%%%%%%%%%%%%%%%%
\section{Introduction}
%%%%%%%%%%%%%%%%%%%%%%%%

Even if a classical system was identical to its "mirror image", its quantum counterpart would not be. This is the essence of the chiral anomaly which appears ubiquitously in all quantum many-body systems involving  massless chiral fermions. The possibility of observing the quantum chiral anomaly at a macroscopic scale has stimulated various explorations and intellectual efforts in several physics research fields from high-energy to condensed matter \cite{Kharzeev:2015znc,Landsteiner:2016led}. Dissipationless transport has been identified as one of the most promising routes in this regard, leading to the discovery of several anomaly-induced phenomena such as the chiral magnetic effect \cite{Fukushima:2008xe}, the chiral separation effect \cite{Metlitski:2005pr} and the chiral vortical effect \cite{Erdmenger:2008rm,Banerjee:2008th}. In addition to these transport properties, the chiral anomaly also manifests itself at macroscopic scale in the low-energy dynamics by originating a novel propagating collective excitation know as the chiral magnetic wave (CMW) \cite{Newman:2005hd,Kharzeev:2010gd}. The CMW represents a propagating wave of fluctuating vector and axial charge densities that is expected to appear, as a result of the interplay between the chiral magnetic effect (CME) and the chiral separation effect (CSE), in chiral anomalous systems when exposed to a background magnetic field $\vec{B}$. 

The experimental identification of propagating CMW in chiral anomalous systems is an active area of research \cite{Ke:2012qb,Belmont:2014lta,STAR:2015wza}. 
In this chase, a prime candidate has always been the quark-gluon plasma (QGP), a highly energetic quantum soup of quarks and gluons generated in heavy ion collisions, as those produced in the relativistic heavy ion collider (RHIC) or large hardron collider (LHC). 
Numerical simulations have suggested that a gapless CMW would induce a chiral dipole moment, resulting in a possibly observable 
charge dependent elliptic flow in QGP \cite{Burnier:2011bf}. 
The observation of a charge dependent elliptic flow has been reported in experiments \cite{Ke:2012qb,Belmont:2014lta,STAR:2015wza,CMS:2017pah}, seemingly supporting the existence of CMW in the QGP. Nevertheless, a thorough analysis indicated that it is impossible to definitively infer the existence of CMW from these experimental signatures \cite{Ke:2012qb,Belmont:2014lta,STAR:2015wza}. This is attributed to the fact that the physical origin of the elliptic flow \cite{Ke:2012qb,Belmont:2014lta,STAR:2015wza} cannot be exclusively ascribed to chiral anomalies, thereby complicating any conclusive statement \cite{CMS:2017pah,Das:2023vbt}. Additionally, from a theoretical perspective, it has been shown that, due to the relaxation of axial charge induced by dynamical gluonic fields and the screening effects associated with the dynamical electromagnetic field (Coulomb interactions), the propagating CMW becomes overdamped and remains therefore elusive to experimental observations \cite{Jimenez-Alba:2014iia,Hattori:2017usa,Shovkovy:2018tks,Landry:2022nog,Grieninger:2023wuq,Grieninger:2023myf,Amoretti:2023hpb}. In \cite{Shovkovy:2018tks}, it was concluded that because of these reasons, it is highly unlikely, if not even impossible, to detect an underdamped CMW in heavy ion collision experiments.

Topological semimetals, such as Dirac or Weyl semimetals (DSM or WSM), represent another important class of chiral anomalous systems \cite{Armitage2018}. Performing measurements on anomalous transport in such systems is certainly simpler compared to the complex QGP high-energy environment. In contrast to QGP, where the axial charge quickly relaxes due to not only the axial anomaly but also the contribution from the dynamical gluons \cite{Gallegos:2018ozs}, the axial charge in DSM or WSM could be nearly conserved, as suggested by the presence of a large negative magnetoresistance \cite{burkov2015negative}. 

Additionally, in topological semimetals, the tunability of the dimensionless parameter $eB/T^2$ (with $B$ the magnetic field, $T$ the temperature and $e$ the electric charge) potentially facilitates the observation of the chiral magnetic wave. Finally, the scattering time of Weyl fermions in topological semimetals spans a large range of values, from the long lifetime Weyl quasiparticle regime \cite{xu2015discovery, lv2015observation} to the hydrodynamic strongly correlated one \cite{Landsteiner:2014vua, Lucas:2016omy,Sukhachov:2018uuz} and plasma regime \cite{gao2016photonic}. All these factors together create favorable conditions that could potentially enable the observation of CMW in such anomalous systems. Therefore, a theoretical study of CMW in topological semimetals is timely and necessary to ascertain this further.

The objective of this study is to investigate the fate of CMW in a class of strongly coupled topological semimetals described by holography \cite{Landsteiner:2015lsa,Landsteiner:2015pdh}, in which time reversal symmetry is broken while parity symmetry is preserved. In a weakly coupled WSM, due to the topological protection of the WSM phase, the system maintains its WSM characteristics even with an increasing separation of the Weyl nodes, provided that no topological phase transition occurs \cite{Landsteiner:2015pdh}. 
Consequently, as the distance between the Weyl nodes in the Brillouin zone increases, the transfer rate of chirality decreases, leading to an almost conserved axial charge. It is natural to expect that similar physics occurs in the holographic counterpart, significantly influencing the dynamics of CMW as already anticipated in \cite{Jimenez-Alba:2014iia}.

On top of that, Coulomb interactions are unavoidable in metals and semimetals \cite{moon2013non,herbut2014mott,janssen2017phase,tchoumakov2019dielectric}, and they undoubtedly influence many of their properties such as the charge susceptibility and the electric conductivity \cite{girvin2019modern}. The chiral magnetic wave, determined by the location of the lowest poles in the density-density correlation functions of electric and axial charge, is also subject to corrections due to Coulomb screening. In the holographic duality, the screening effect is attributed to the presence of a dynamical gauge field in the boundary system, achievable by using mixed-boundary conditions for the bulk gauge fields or equivalently the double-trace deformation (DTD) method \cite{Witten:2001ua,Berkooz:2002ug,Marolf:2006nd}. This approach has been used in the literature in many instances, \textit{e.g.}, \cite{Domenech:2010nf,Montull:2011im, %Montull:2012fy,Salvio:2012at,Salvio:2013jia,
gran2020plasmons,mauri2019screening,Ahn:2022azl,Jeong:2023las,Baggioli:2023oxa,PhysRevD.106.086005}. 

The focus of this work is to study the nature of the CMW, and the feasibility of its experimental observation, in strongly coupled Weyl semimetals by taking into account the effects of both axial charge relaxation and Coulomb interactions. More in general, our investigation aims to offer theoretical insights for potential experimental setups in table-top condensed matter systems.

This paper is organized as follows. In Section \ref{sec:hydro}, we construct the hydrodynamic theory for chiral magnetic waves with and without Coulomb interactions. In Section \ref{sec:holomodel}, we review the holographic Weyl semimetal model and present the bulk solutions in the presence of a magnetic field in the probe limit. In Section \ref{Sec:screening}, we introduce Coulomb interactions in the holographic Weyl semimetals using mixed boundary conditions. In Section \ref{sec:CMW}, we investigate the dispersion relation of the collective chiral magnetic waves in strongly coupled Weyl semimetals by using the aforementioned holographic model. We conclude and discuss the most relevant open questions in Section \ref{sec:conclusion}. The computational details and additional information can be found in Appendices \ref{App:A} to \ref{app:screening}.

\section{Hydrodynamics of chiral magnetic waves in Dirac and Weyl semimetals}
\label{sec:hydro}

Chiral magnetic waves appear as a direct consequence of the combination of the chiral magnetic effect (CME) and the chiral separation effect (CSE) in the presence of a background magnetic field. As a result of these two effects, the vector charge $\rho_V$ and the axial charge $\rho_A$ are able to generate each other inducing the propagation of a pair of longitudinal waves -- chiral magnetic waves. Aside from their longitudinal nature, this phenomenon is very similar to the propagation of electromagnetic waves created by the interplay between magnetic and electric fields and described by standard Maxwell equations. Recently, important analogies have been also found in comparison to the emergence of second sound in superfluids \cite{Delacretaz:2019brr}.

The chiral magnetic effect refers to the generation of a vector current along a background magnetic field direction $B$ induced by a finite axial charge density \cite{Fukushima:2008xe}, \textit{i.e.}, 
\bea
J_V^z=\sigma_{\text{CME}}B_z\,,\quad\quad
\sigma_{\text{CME}}=8\alpha\mu_A\,,
\eea
where, without loss of generality, we choose the magnetic field along the $z$ direction. Here, $\alpha$ is the chiral anomaly coefficient, $\sigma_{\text{CME}}$ the chiral magnetic conductivity, and $\mu_A$ the axial chemical potential that relates to the axial charge density $\rho_A=\chi_{A} \mu_A$ via the corresponding axial charge susceptibility $\chi_{A}$.  Similarly, the chiral separation effect is the generation of axial current along a background magnetic field direction $B_z$ induced by a finite charge density, \textit{i.e.},
\bea
J_A^z=\sigma_{\text{CSE}}B_z\,,\quad\quad
\sigma_{\text{CSE}}=8\alpha\mu_V\,,
\eea
where $\mu_V$ is the vector chemical potential. The vector charge density is then given by $\rho_V=\chi_{V} \mu_V$ with $\chi_{V}$ being the vector charge susceptibility.

The expressions mentioned above, and defining the CME and CSE, are valid only in the homogeneous case, where all quantities do not depend on the spatial coordinates. After introducing spatial dependence along the coordinate $z$, and taking into consideration dissipative dynamics, the constitutive equations for the vector and axial currents are given, to first order in derivatives $O(\partial)$ \cite{Kharzeev:2010gd}, by
\bea
\begin{split}
J_V^z&=\sigma_{\text{CME}}B-\sigma_V\partial_z\mu_V
=\frac{8\alpha B}{\chi_A}\rho_A-D_V\partial_z\rho_V\,,\\
J_A^z&=\sigma_{\text{CSE}}B-\sigma_A\partial_z\mu_A
=\frac{8\alpha B}{\chi_V}\rho_V-D_A\partial_z\rho_A\,,
\end{split}
\label{eq:constitutive}
\eea
where $B=B_z$. In the constitutive equations, Eq.\eqref{eq:constitutive}, we have assumed that parity symmetry is preserved and only time reversal symmetry is broken, to have the same symmetry pattern with the holographic Weyl semimetal set-up in Sec.\ref{sec:holomodel}. $\sigma_V, \sigma_A$ are respectively the vector and axial conductivities, and 
\bea
D_V=\frac{\sigma_V}{\chi_V}\,,\quad\quad\quad
D_A=\frac{\sigma_A}{\chi_A}\,,
\eea
are the diffusion constants for the vector and axial charges, respectively (Einstein relations). Moreover, in Eq.\eqref{eq:constitutive} we have used the concrete expression for the CME and CSE conductivities from \cite{Fukushima:2008xe, Metlitski:2005pr}. %\MB{add} 
Note that the susceptibilities $\chi_{A,V}$ and conductivities $\sigma_{A,V}$ in the constitutive equations Eq.\eqref{eq:constitutive} are just one contribution to the total DC conductivities once axial charge relaxes over a long time scale introducing a Drude-like term, see App.\ref{App:B}. When the axial charge is conserved, the first order hydrodynamic transport coefficients in Eq.\eqref{eq:constitutive} directly enter in the real part of the total DC conductivities. 

In presence of a magnetic field $\vec{B}$ along $z$ direction, but in absence of an electric field $\vec{E}$, one has $\vec{E}\cdot \vec{B}=0$ and the axial anomaly does not contribute to any dynamics. However, the U(1) axial symmetry is not a fundamental symmetry of nature; hence, axial charge is not a conserved quantity but it rather decays at a rate $\Gamma \equiv 1/\tau_5$. In Weyl semimetals, the  axial charge relaxation time $\tau_5$ is usually much larger than the quasi-particles lifetime \cite{zhang2016signatures}. Therefore, axial charge is approximately conserved. The dynamics are then governed by the conservation of the U(1) vector current and the softly broken non-conservation of the axial one \cite{Jimenez-Alba:2014iia},
\bea
\partial_\mu J^\mu_V=0\,,\quad\quad\quad
\partial_\mu J^\mu_A=-\Gamma\rho_A\,.
\label{eq:hydrolaw}
\eea
Here, the axial charge relaxation rate $\Gamma$ is a phenomenological quantity that depends on temperature, magnetic field and the separation between the Weyl nodes. We notice that the second equation in Eqs.\eqref{eq:hydrolaw} is meaningful only in the limit in which $\Gamma$ is small (compared to the characteristic timescale of the system).

Combining the constitutive equations Eqs.\eqref{eq:constitutive} with  Eqs.\eqref{eq:hydrolaw}, we obtain the dispersion relations for the two lowest excitations
\bea
\omega_{\pm}=-\frac{i\Gamma}{2}-\frac{i}{2}(D_A+D_V)k_z^2\pm\sqrt{\frac{(8\alpha B)^2}{\chi_A\chi_V}k_z^2-\frac{\left((D_A-D_V)k_z^2+\Gamma\right)^2}{4}}\,.
\label{eq:dispersion}
\eea
In the limit of small breaking of axial charge conservation, eventual corrections to $\sigma_A, \chi_A, D_A$ can be ignored, and one can consistently assume for simplicity that $\chi_A=\chi_V=\chi, \sigma_A=\sigma_V=\sigma$ and that all parameters only depend on $B$ (see \cite{Kharzeev:2010gd} for more details about the validity of this limit). Then, Eq.\eqref{eq:dispersion} can be simplified to \cite{Jimenez-Alba:2014iia} 
\bea
\omega_{\pm}=-\frac{i\Gamma}{2}-iDk_z^2\pm\frac{1}{2}\sqrt{\frac{(16\alpha B k_z)^2}{\chi^2}-\Gamma^2}\,.
\label{eq:CMWtype1}
\eea
The above dispersion relation indicates that, in presence of axial charge relaxation, chiral magnetic waves are not gapless propagating sound waves anymore. On the contrary, the real part of their dispersion is nonzero only above a critical wave-vector. This type of dispersion, gapped in momentum space, appears in several branches of physics \cite{BAGGIOLI20201}. As a concrete example, it is the dispersion expected for shear waves in classical liquids \cite{PhysRevLett.118.215502}.

\vspace{.3cm}
In addition to axial charge relaxation, in Dirac and Weyl semimetals, the inevitable Coulomb interactions play an important role as well. For example, at zero doping, Coulomb interactions can induce non-Fermi liquid behavior \cite{moon2013non}, and a further phase transition into an interaction-driven topological insulator \cite{herbut2014mott,janssen2017phase}. 
When we consider chiral magnetic waves, Coulomb interactions are also non-negligible. For example, the dielectric function gets screened by Coulomb effects \cite{tchoumakov2019dielectric}. The important effects of Coulomb interactions on chiral magnetic waves have been noticed and discussed for the first time in \cite{Shovkovy:2018tks}, casting serious doubts about the potential experimental detection of the latter in the quark-gluon plasma.

Simplifying the analysis by restricting it along the $z$ direction, the parity-preserved longitudinal currents can be now written as
\bea
\begin{split}
J^V_z&=\sigma_{\text{CME}}\,B+\sigma\left(E_z-\partial_z\mu_V\right)
=\frac{8\alpha B}{\chi_A}\rho_A-D\,\partial_z\rho_V+\sigma E_z\,,\\
J^A_z&=\sigma_{\text{CSE}}\,B-\sigma\partial_z\mu_A
=\frac{8\alpha B}{\chi_V}\rho_V-D\partial_z\rho_A\,
\end{split}
\label{constitutive}
\eea
where, as a consequence of dynamical electromagnetism, an electric field is naturally generated by the presence of charge. In particular, the electric field satisfies Gauss's law $\vec{\nabla}\cdot\vec{E}=\rho_V/\varepsilon_e$ where $\varepsilon_e$ is the electric permittivity. Let us emphasize that, at this point, the electric field $E_z$ is not external but rather a dynamical field. Eqs. \eqref{constitutive} are the most general constitutive relations for the currents, since there are no Hall effect terms for the longitudinal currents in the probe limit, \textit{i.e.}, when we ignore the variations of the temperature $\delta T$ and velocity $\delta u^\mu$. Also, terms like $B\cdot\delta\sigma_{\text{CME}}$ and $B\cdot\delta\sigma_{\text{CSE}}$, that capture non-linear effects in the magnetic field, do not appear at this order. As shown below, in our setup, the dynamical magnetic field vanishes and, as a consequence, there are no terms like $\sigma_{\text{CME}}\delta B$ or $\sigma_{\text{CME}}\delta B$ either. 

The above analysis is consistent with Maxwell equations, 
\bea
\label{eq:Maxwelleqn}
\begin{split}
\vec{\nabla}\cdot \vec{E}&=\frac{\rho_V}{\varepsilon_e}\,,\\
\vec{\nabla}\cdot\vec{B}&=0\,,\\
\vec{\nabla}\times \vec{B}&=\mu_m \vec{J}+\varepsilon_e\mu_m\frac{\partial\vec{E}}{\partial t}\,,\\
\vec{\nabla}\times\vec{E}&=-\frac{\partial\vec{B}}{\partial t}\,
\end{split}
\eea
where $\mu_m$ is the magnetic permeability. 
In the presence of a charge density $\rho_V(t,z)$, a longitudinal electric field $E_z(t,z)$ is generated according to Gauss's law. Since $\vec{\nabla}\times\vec{E}=0$, there is no time-dependent magnetic field generated by the dynamical electric field, and we have only an external background magnetic field $B_z$. Finally, $\vec{\nabla}\cdot\vec{B}=0$ is trivially satisfied and Ampere's law reduces to $ \vec{J}+\varepsilon_e\frac{\partial\vec{E}}{\partial t}=0$. This equation is nothing but the relaxation of electric lines in the presence of charged matter, which corresponds to the explicit breaking of the electric $1$-form global symmetry \cite{Grozdanov:2016tdf}.

Because of the EM-generated dynamical electric field, the axial anomaly does not vanish anymore and the dynamical equations become
\bea
\partial_\mu J^\mu_V=0\,,\quad\quad\quad
\partial_\mu J^\mu_A=8\alpha~ \vec{E}\cdot \vec{B}-\Gamma\rho_A\,.
\label{conservation}
\eea
The above equations, together with the constitutive equations, lead to the corresponding dispersion relation for chiral magnetic waves,
\bea
\omega_{\pm}=-\frac{i\Gamma}{2}-\frac{i\sigma}{2\varepsilon_e}-iDk_z^2\pm\frac{\sqrt{4\varepsilon_e\left(8\alpha B\right)^2\left(\chi+\varepsilon_e k_z^2\right)-(\sigma-\varepsilon_e\Gamma)^2\chi^2}}{2\varepsilon_e\chi}\,.
\label{eq:CMWtype3}
\eea
Notice that the dispersion relation Eq.\eqref{eq:CMWtype3} gets corrected by both axial relaxation and Coulomb screening effects, which is beyond the results of \cite{Kharzeev:2010gd}, where neither of these effects is included, and of \cite{Jimenez-Alba:2014iia} where only axial relaxation is considered. 

Using a non-relativistic hydrodynamic framework for WSM, CMW along with other collective modes have been investigated in \cite{Sukhachov:2018uuz}, where a steady state of the chiral electron fluid is realized by introducing momentum dissipation in the Euler equation. In contrast, our setup does not consider the dynamics of energy and momentum (which are not relevant for the holographic model in the probe limit) and Ohm's law appears directly in the constitutive equation for the dynamical electric field self-generated by the near-equilibrium Weyl plasma. Let us stress that the analysis in \cite{Sukhachov:2018uuz} is limited to the hydrodynamic framework where the hydrodynamic parameters cannot be determined. On the contrary, in our case we will make use of the holographic model as a microscopic description allowing a direct estimate of all the hydrodynamic coefficients as well.

The results in absence of Coulomb interactions presented above in Eq.\eqref{eq:CMWtype1} can be obtained by sending $\varepsilon_e \rightarrow \infty$ in Eq. \eqref{eq:CMWtype3}. We remind that $\chi$ and $\sigma$ here are the unscreened transport coefficients computed in absence of Coulomb interactions. Later on, we will discuss this point in more detail, and we will show that the dispersion relations in Eqs.\eqref{eq:CMWtype3} are nothing but the poles of the screened current-current correlator.

%%%%%%%%%%%%%%%%%%%%%%%%%%%%%%%%%%%%%%%%%%%%%%%%%%%%%%%%%%
\section{Magnetic field effects on holographic Weyl semimetals}
%%%%%%%%%%%%%%%%%%%%%%%%%%%%%%%%%%%%%%%%%%%%%%%%%%%%%%%%%%
\label{sec:holomodel}
We consider the holographic Weyl semimetal model \cite{Landsteiner:2015lsa,Landsteiner:2015pdh} (see \cite{Landsteiner:2019kxb} for a review) described by the five-dimensional bulk action
\begin{align}
\label{eq:holomodel}
  S=&\int d^5x \sqrt{-g}\bigg[\frac{1}{2\kappa^2}\Big(R+\frac{12}{L^2}\Big)-\frac{1}{4e^2}\mathcal{F}^2-\frac{1}{4e^2}F^2  -(D_\mu\Phi)^*(D^\mu\Phi)-V(\Phi)  \\
 &+\epsilon^{\mu\nu\rho\sigma\tau}A_\mu\bigg(\frac{\alpha}{3} \Big(F_{\nu\rho} F_{\sigma\tau}+3 \mathcal{F}_{\nu\rho}  \mathcal{F}_{\sigma\tau}\Big) \bigg)\bigg]\,,\nonumber
\end{align} 
with $\mathcal{F}_{\mu\nu}=\partial_\mu V_\nu-\partial_\nu V_\mu, F_{\mu\nu}=\partial_\mu A_\nu-\partial_\nu A_\mu$ and $D_\mu\Phi = (\partial_\mu - i q A_\mu)\Phi$. The axial anomaly in the dual field theory is introduced using Chern-Simons terms proportional to $\alpha$ that are determined by the anomalous structure of the currents. The ratio between the coefficients of $A\wedge F\wedge F$ and $A\wedge \mathcal{F}\wedge \mathcal{F}$ is fixed by considering the consistent current choice where the vector current is conserved and only the conservation of axial current is broken (by anomaly).\footnote{We use the following convention for Levi-Civita symbol $\tilde{\epsilon}_{txyzr}=1$ or $\tilde{\epsilon}^{txyzr}=-1$, and the Levi-Civita tensor $\epsilon_{\mu\nu\rho\sigma\tau}\equiv\sqrt{-g}\tilde{\epsilon}_{\mu\nu\rho\sigma\tau}$ and $\epsilon^{\mu\nu\rho\sigma\tau}\equiv\frac{1}{\sqrt{-g}}\tilde{\epsilon}^{\mu\nu\rho\sigma\tau}$.} The complex %real 
scalar field $\Phi$ couples to the axial vector field $A_\mu$ and its scalar potential is given by $V(\Phi)=m^2|\Phi|^2$. Whenever this scalar has a non-zero boundary source, axial symmetry is explicitly broken and axial charge relaxes. Here, we choose the scalar mass $m^2=-3$ and switch on only the $z$ component of the axial gauge field $A_z$. Without loss of generality, we will choose $2\kappa^2=L^2=e^2=1$ in the rest of the manuscript.  

With these definitions, the consistent currents in the dual field theory can be defined by the following expressions \cite{Landsteiner:2016led}
\bea
J_V^\mu&=\displaystyle{\lim_{r\to\infty}}\frac{\delta S}{\delta V_\mu}=\displaystyle{\lim_{r\to\infty}}\sqrt{-g}\left(\mathcal{F}^{\mu r}+4\alpha \epsilon^{r\mu\nu\rho\sigma}A_\nu\mathcal{F}_{\rho\sigma}\right)
\label{eq:JV}
\,,\\
J_A^\mu&=\displaystyle{\lim_{r\to\infty}}\frac{\delta S}{\delta A_\mu}=\displaystyle{\lim_{r\to\infty}}\sqrt{-g}\left(F^{\mu r}+\frac{4}{3}\alpha \epsilon^{r\mu\nu\rho\sigma}A_\nu F_{\rho\sigma}\right)\,,
\label{eq:JA}
\eea
where $r$ is the radial coordinate in the bulk and $r=\infty$ is the UV AdS boundary. 

The Ward identities for the consistent currents are then
\bea
\begin{split}
\partial_\mu J^\mu_V&=0\,,\\
\partial_\mu J^\mu_A&=\displaystyle{\lim_{r\to\infty}}\sqrt{-g}\left(-\frac{\alpha}{3}\epsilon^{r\mu\nu\rho\sigma}\left(F_{\mu\nu}F_{\rho\sigma}+3\mathcal{F}_{\mu\nu}\mathcal{F}_{\rho\sigma}\right)-iq[\Phi(D^r\Phi)^*-\Phi^*(D^r\Phi)]\right)+\text{c.t.}\,,
\end{split}
\eea
where ``c.t.'' stands for contact terms. As expected, the vector current is conserved. On the contrary, the axial current is not conserved due to the axial anomaly proportional to $\alpha$ and the effects of the scalar field $\Phi$. When $\vec{E}=0\,,\vec{B}\neq0$, only a source for the scalar field can explicitly break axial charge conservation. 

\subsection{The solution in the probe limit}

We work in the probe limit in the presence of background magnetic field and at finite temperature. In the context of holographic Weyl semimetals, the probe limit is a reasonable approximation as long as the system is not too close to the quantum critical point and $B_z$ is not parametrically larger than temperature \cite{Jimenez-Alba:2015awa}. 

In order to study chiral magnetic waves in WSM, we consider the simplest setup with 
\bea
\mu=\mu_5=\rho_5=0\,,\quad \rho\neq 0\,,\quad \vec{E}=0\,,\quad B_z\neq 0\,,\quad b_z\neq 0\,,\quad M\neq 0\,.
\eea
From the bulk point of view, this corresponds to the AdS-Schwarzschild solution where we neglect the backreation of matter field on the metric, \textit{i.e.}, 
\bea
\begin{split}
ds^2&=-udt^2+\frac{dr^2}{u}+r^2d\vec{x}^2\,,\quad u(r)=r^2\left(1-\frac{r_h^4}{r^4}\right)\,,\\
V&=V_t dt+\frac{B}{2}(ydx-xdy)\,,\quad
A=A_zdz\,,\quad \Phi=\phi(r)\,.
\end{split}
\eea
Here $B=B_z$ is in the same direction of $z$. 
 
The asymptotic behaviors of the matter fields near to the AdS boundary, $r\rightarrow \infty$, are given by
\bea
\begin{split}
r\phi(r)&=M+...\,,\\
A_z(r)&=b+...\,,\\
V_t(r)&=\mu+...\,.
\end{split}
\eea
where $M$ corresponds to a source for the dual scalar operator that breaks U(1) axial symmetry, $b$ relates to the separation of the chiral nodes and $\mu$ is the chemical potential in the boundary field theory. We will be interested in zero chemical potential states. 

The equations of motion for $A_z\,, \phi\,,$ and $V_t$ are
\bea
\begin{split}
A_z''+\left(\frac{1}{r}+\frac{u'}{u}\right)A_z'-\frac{2q^2\phi^2}{u}A_z+\frac{8\alpha B}{ru}V_t'&=0\,,\\
\phi''+\left(\frac{3}{r}+\frac{u'}{u}\right)\phi'-\left(\frac{m^2}{u}+\frac{q^2A_z^2}{r^2u}+\frac{\lambda\phi^2}{u}\right)\phi&=0\,,\\
V_t''+\frac{3}{r}V_t'+\frac{8\alpha B}{r^3}A_z'&=0\,. 
\end{split}
\label{eom:probe}
\eea
Notice that the last equation in \eqref{eom:probe} reduces to a total derivative 
\bea
(r^3V_t'+8\alpha B A_z)'=0\,,
\eea
and its solution can be formally written as
\bea
V_t(r)=\int_{r_0}^r \frac{ds}{s^3}\Big(c_1-8\alpha B A_z(s)\Big)+c_2\,,
\eea
where $c_1$ and $c_2$ are integration constants.
As required by the regularity conditions, $V_t$ vanishes at the horizon, \textit{i.e.}, $c_2=0$. 
The choice of $c_1$ is arbitrary and is related to the chemical potential $\mu$ in the field theory. We will focus on zero chemical potential states $\mu=0$ via choosing $c_1$ that is a function of the parameters in the boundary field theory. 

For later convenience, we define the dimensionless parameters in our system as
\bea
\tilde{B}\equiv\frac{B}{\pi^2 T^2}\,,\quad\quad
\tilde{M}\equiv\frac{M}{\pi T}\,,\quad\quad
\tilde{b}\equiv\frac{b}{\pi T}\,.
\eea

\subsection{Magnetic field effects on the phase diagram}

An interesting question arises when we switch on a magnetic field in a Weyl semimetal, that is, does the topological phase still exist or is it broken by the magnetic field? In other words, does the magnetic field change the phase diagram and the topological phase transition between a topological Weyl semimetal to a gapped trivial phase? 

In the weakly coupled field theory \cite{Li:2016czu,Bruni:2023ife}, this can be investigated by solving the Landau levels from the Hamiltonian where the electrons couple to the magnetic field $B$. One can find \cite{Li:2016czu,Bruni:2023ife} that the chiral conduction band splits into an infinite number of Landau levels in the presence of $B$. Importantly, the lowest Landau level remains the chiral band and it is unchanged by $B$. More precisely, via tuning the ratio $M/b$ between the mass and the time-reversal breaking in the Lorentz violating Dirac equation \cite{Colladay:1998fq}
\bea
\left[\gamma^\mu\left(i\partial_\mu-eV_\mu\right)-\gamma^5\gamma^zb-M\right]\Psi=0
\eea
in the presence of the magnetic field, $V_\mu=(0,By/2,-Bx/2,0)$, the topological phase transition pattern remains unchanged. Now, the energy spectrum is made of Landau Levels. The lowest one is defined by the dispersion 
\be
E_0=b\pm\sqrt{k_z^2+M^2}\,,
\label{LLL}
\ee
where we have set $k_x=k_y=0$. Expression \eqref{LLL} takes exactly the same form as without magnetic field. The detailed expression for higher Landau levels can be found in \textit{e.g.} \cite{Bruni:2023ife}. The key observation is that the magnetic field does not alter the lowest band structure. 
However, this analysis is reliable only when the magnetic field is sufficiently small, since otherwise the degrees of freedom from the higher Landau levels would get excited as well, spoiling this approximation. 

In strongly coupled Weyl semimetals, the notion of electronic bands is absent and 
the anomalous Hall conductivity (AHE) serves as the order parameter to identify the Weyl semimetal phase and the topological phase transition \cite{Landsteiner:2015lsa,Landsteiner:2015pdh}.  
To compute the AHE, the transverse sector of fluctuations at $k_z=0$ is switched on, \textit{i.e.}, $\delta v_x, \delta v_y, \delta a_x, \delta a_y$. The EOMs for the transverse sector are summarized in Appendix \ref{app:transverse}. It turns out that in the probe limit with background magnetic field the holographic formula for the anomalous Hall effect remains unchanged, 
\bea
\label{eq:ahe}
\sigma_{\text{AHE}}=8\alpha A_z(r_h)\,,
\eea
and depends only on the horizon behavior of $A_z$. 

\begin{figure}[H]
\centering
\includegraphics[width=0.55\textwidth]{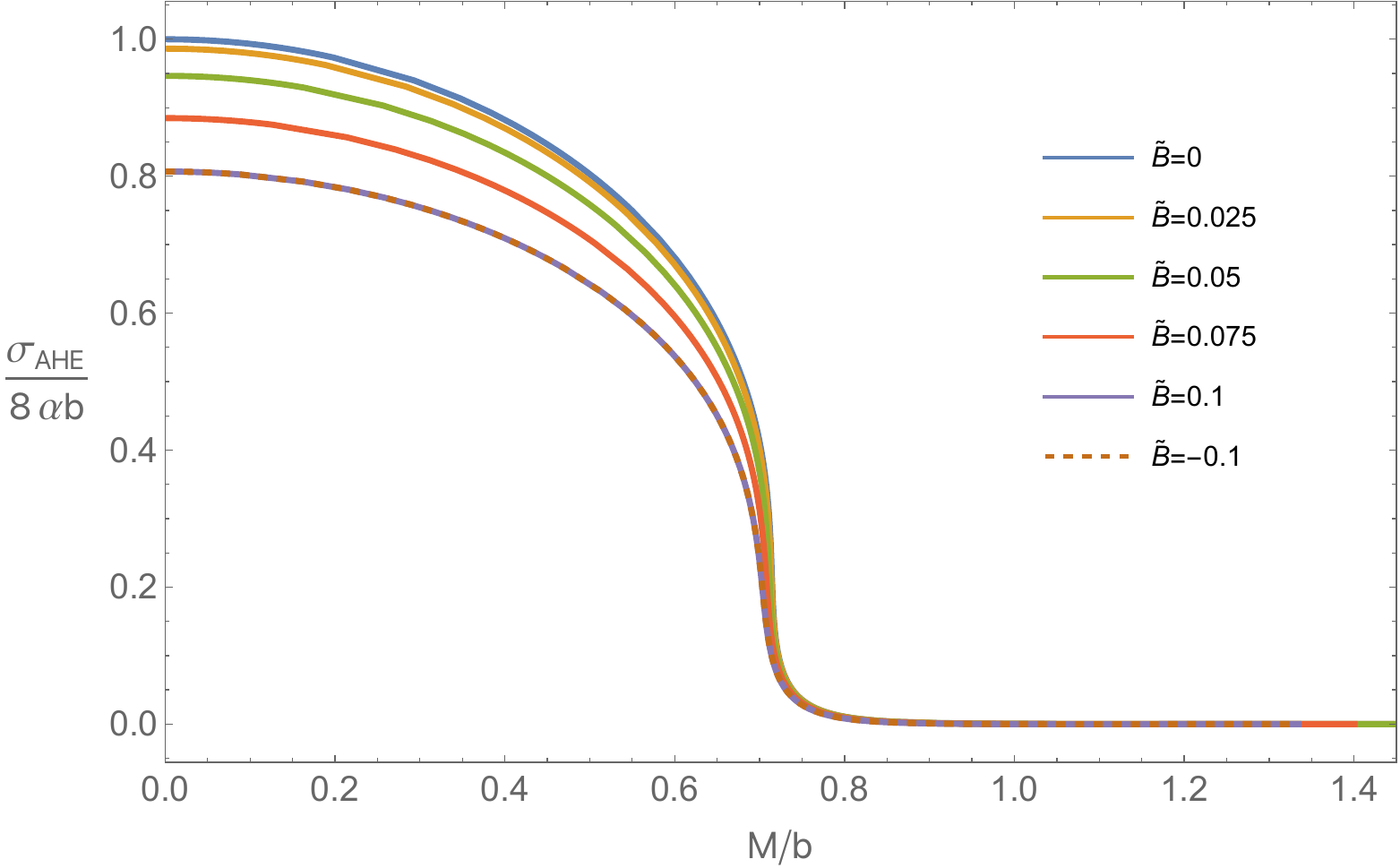}
\caption{Anomalous Hall conductivity $\sigma_{\text{AHE}}$ as a function of $M/b$ for different, and small values of the dimensionless magnetic field $\tilde B$, at a low temperature $T/b=1/(8\pi)$}.
\label{fig:AHE}
\end{figure}

The structure of the phase diagram for different values of the magnetic field $B$ can be obtained  by looking at the order parameter $\sigma_{\text{AHE}}$ shown in Fig.\ref{fig:AHE}. For small magnetic  field, for which the probe limit is valid\footnote{The backreaction of the magnetic field might have nontrivial effects on magnetotransport properties, \textit{e.g.}, \cite{Sun:2016gpy}.}, the inflection point does not move. Hence, the location of the quantum critical point in terms of the dimensionless parameter $M/b$ does not depend on $B$ (or on the sign of $B$). Interestingly, this phenomenon is the exactly same as in the weakly coupled theory in which the Weyl semimetal phase and the topological phase transition are dominated by the lowest Landau level that is independent of $B$.

In the backreacted case with finite magnetic field, 
we do not have such a simple expression for AHE as Eq.\eqref{eq:ahe}. This can be seen as follows. First, let us notice that there exists a topological Chern-Simons charge induced by the anomaly, \textit{i.e.}, $\rho_{CS}\propto \vec{b}\cdot\vec{B}$. In the  gravitation theory, this indicates the necessity of having a nonzero time component of the vector gauge field $V_t\neq 0$. When we switch on the fluctuations to compute anomalous Hall conductivity, this non-vanishing potential couples the fluctuations of vector and metric fields together, making the computations complicated. In this case the anisotropy of the butterfly velocity has been investigated as an effective probe of the phase transition \cite{Bruni:2023ife}. In this work, we focus on the probe limit for simplicity. 

\subsection{Axial charge relaxation time}
\label{axial relaxation time}

When axial charge conservation is weakly broken, the conservation equation is modified 
into
\bea
\partial_\mu J^\mu_A=-\Gamma \rho_A\,,
\eea
which indicates that the axial charge $\rho_A$ decays with a relaxation time $\tau_5\equiv 1/\Gamma$. This relaxation time $\tau_5$ can be computed by looking at the corresponding purely imaginary pole in the axial charge density-density correlation function of, \textit{i.e.}, 
\bea
\tau_5^{-1}\equiv -\text{Im}(\omega^*)=\Gamma\,,
\eea 
where $\omega^*$ is the non-hydrodynamic pole related to axial charge relaxation. In the holographic dual picture, the pole in the correlation function can be derived from the quasi-normal modes (QNM) spectrum of the bulk theory.  

\begin{figure}[H]
\centering
\includegraphics[width=0.49\textwidth]{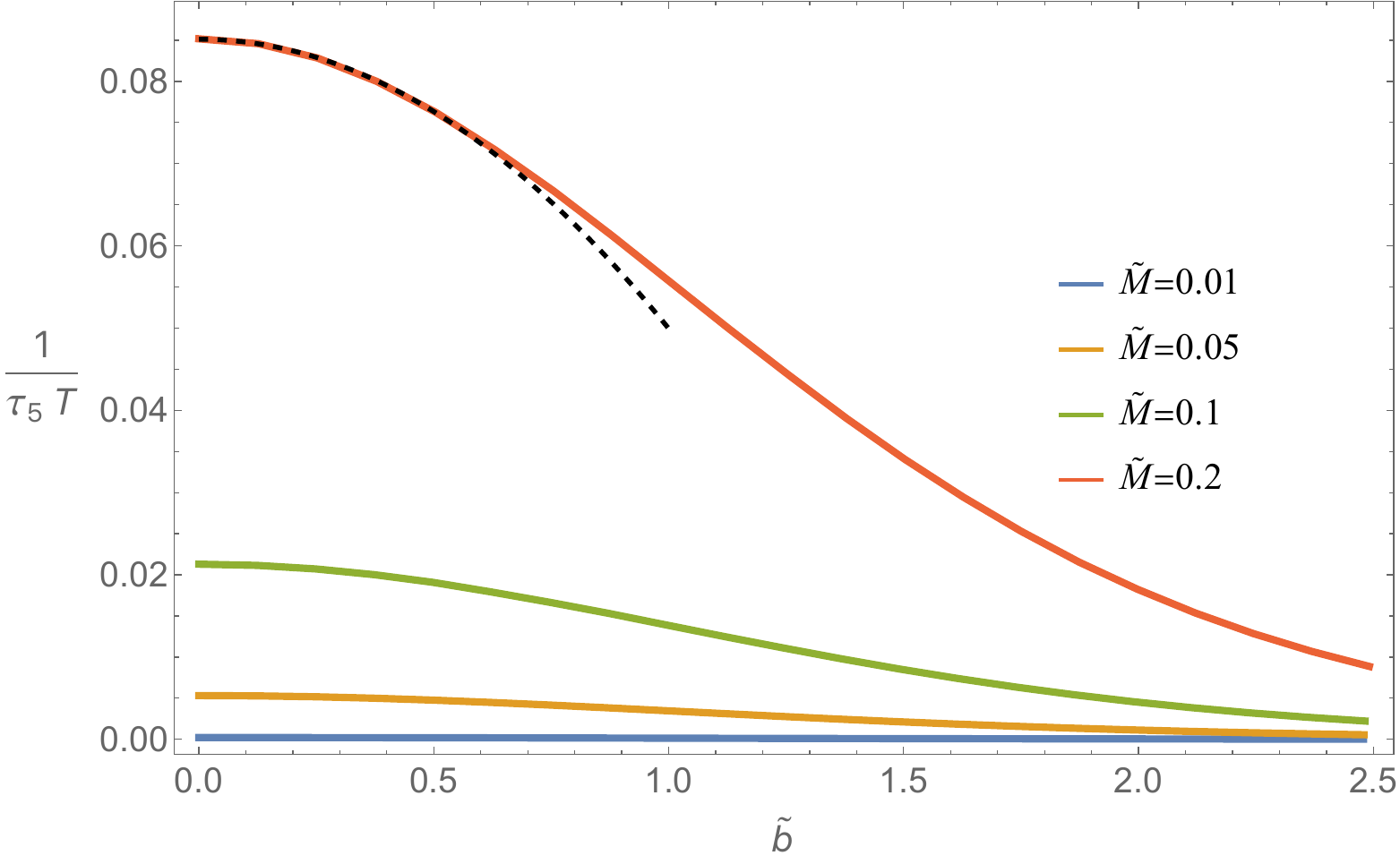}
\includegraphics[width=0.49\textwidth]{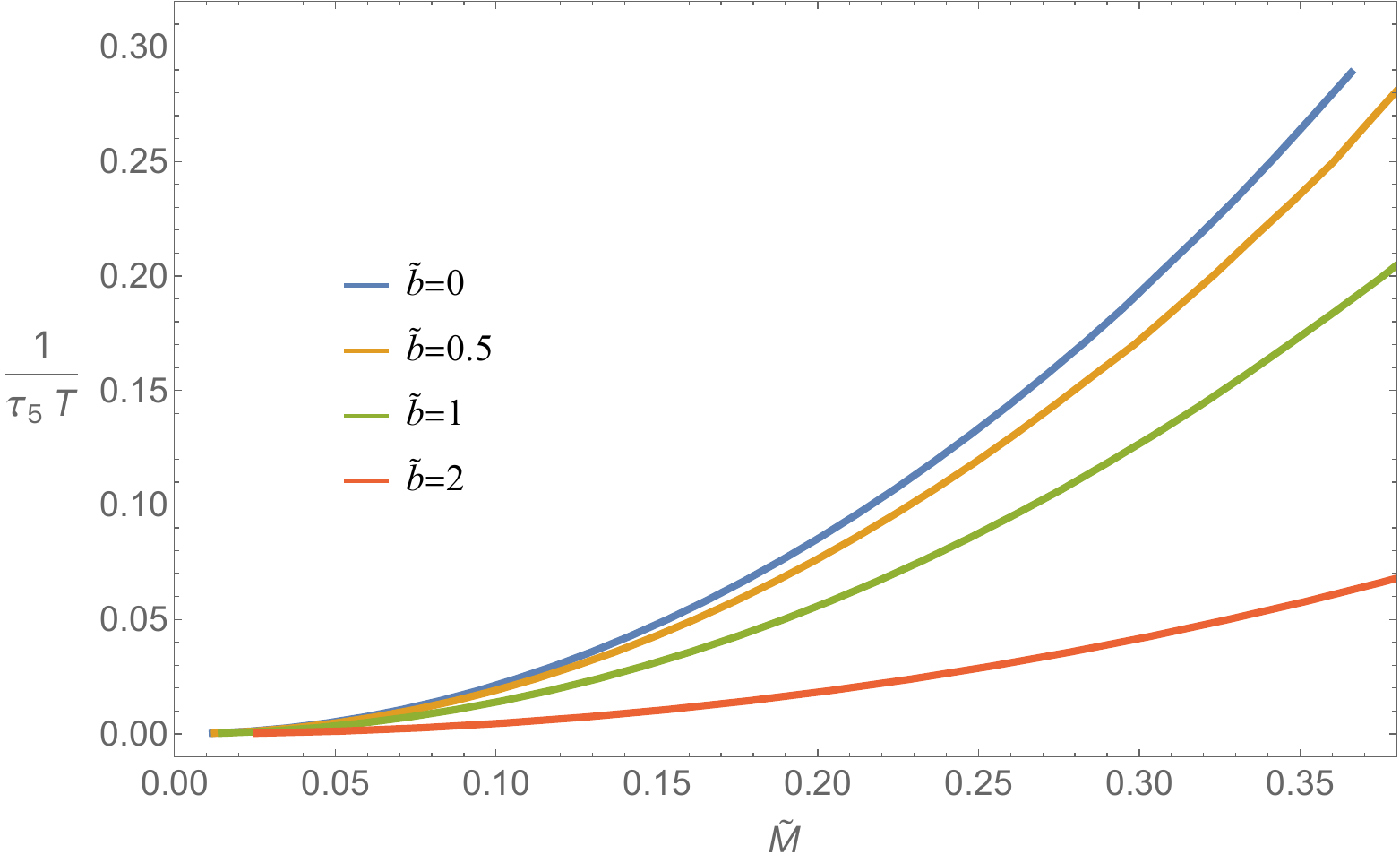}
\caption{{\bf Left:} The dimensionless axial relaxation time as a function of $\tilde{b}$ for different values of $\tilde{M}$ with fixed $\tilde{B}=0.1$. In the small $\tilde{b}$ regime, the decrease in $(\tau_5 T)^{-1}$ obeys a square law in terms of $\tilde{b}$ which is shown by the dashed black curve. {\bf Right:} $(\tau_5 T)^{-1}$ as a function of $\tilde{M}$, for different values of $\tilde{b}$.}
\label{fig:tauVSb}
\end{figure}

The dependence of the axial relaxation time $\tau_5$ on $\tilde b$ is shown in the left panel of Fig.\ref{fig:tauVSb}. The horizontal axis represents the dimensionless separation of the Weyl nodes in wave-vector space, $\tilde{b}$, while the vertical axis represents the inverse axial charge relaxation time normalized by temperature, $(\tau_5 T)^{-1}$. We observe that $(\tau_5 T)^{-1}$ decreases monotonically by increasing $\tilde{b}$ for all values of $\tilde{M}$. In the small $\tilde{b}$ regime, we find that the decay can be well approximated by a quadratic power law scaling. In the large $\tilde{b}$ regime,  $\tau_5^{-1}$ decays to zero. This indicates that even in the small external magnetic field regime, $B\ll T^2$, with fixed $\tilde{M}$, $\tau_5 T$ can be made arbitrarily large by taking the limit $M/b\rightarrow 0$. This is consistent with the physical intuition that the larger the distance between the two chiral nodes in the Brillouin zone, the more inhibited the processes between the two nodes that do not conserve chirality. For completeness, we show the dependence of $(\tau_5T)^{-1}$ on $\tilde{M}$ for different $\tilde{b}$ in the right panel of Fig.\ref{fig:tauVSb}. The results are consistent and show that the axial relaxation rate $\Gamma$ follows approximately a quadratic power law behavior $\Gamma \simeq a  \tilde{M}^2$ where the coefficient $a$ decreases as we increase $\tilde{b}$. 

\begin{figure}[H]
\centering
\includegraphics[width=0.49\textwidth]{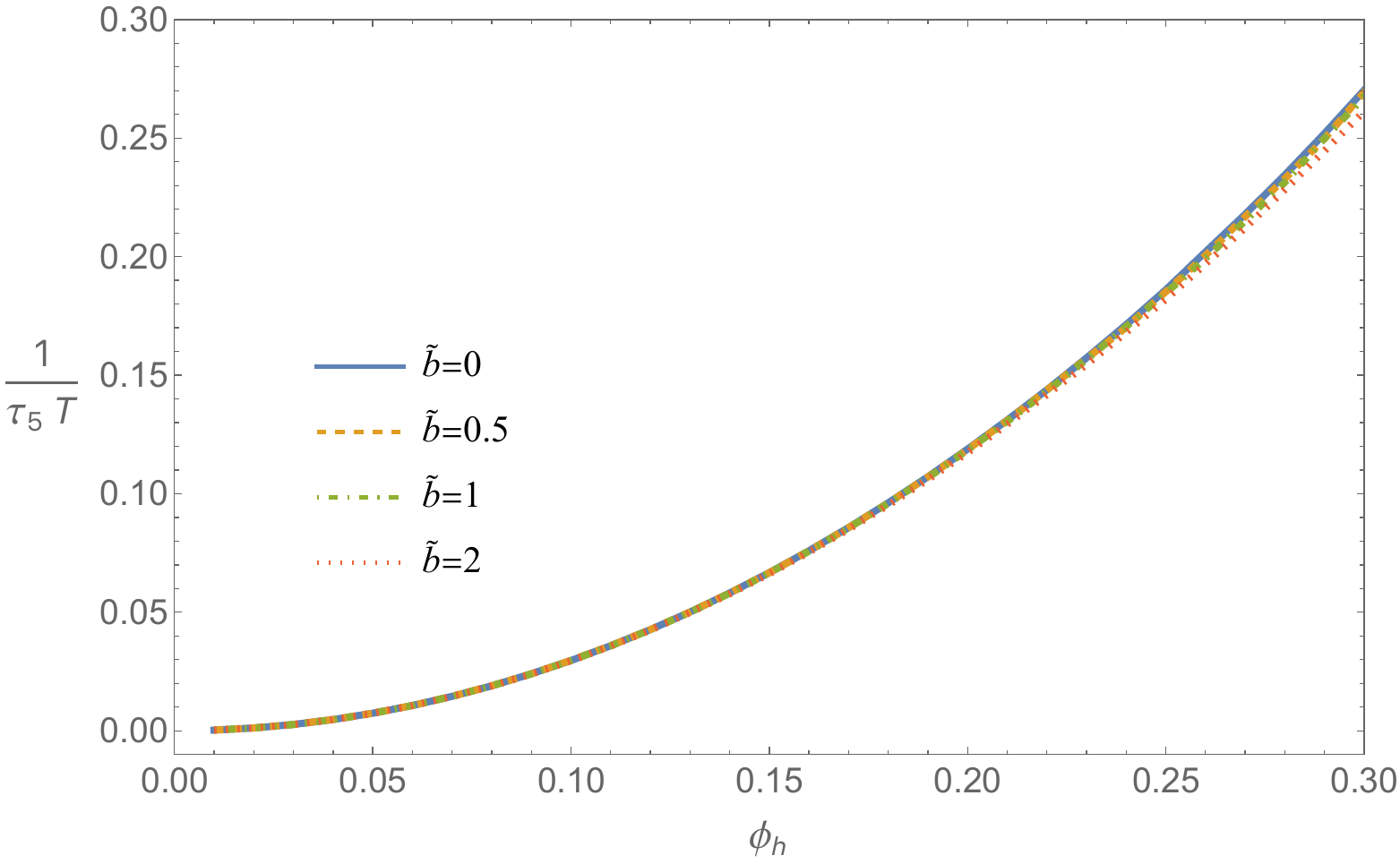}
\includegraphics[width=0.49\textwidth]{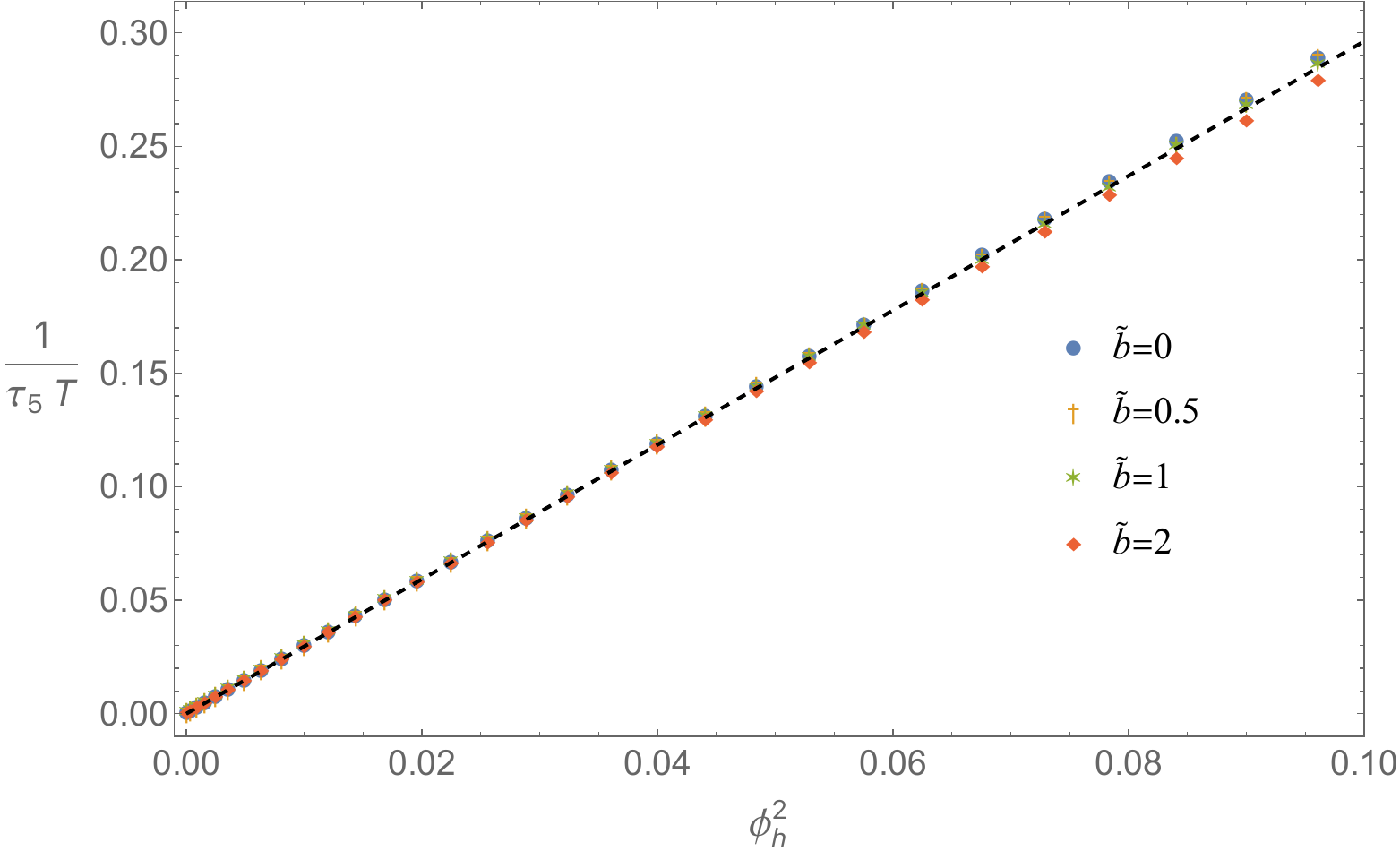}
\caption{The universal relation between $\Gamma$ and the horizon value of the bulk scalar field, \textit{i.e.} $\phi_h\equiv \phi(r_h)$. In the right panel, a different representation which emphasizes the universal behavior $\tau_5^{-1} \simeq \frac{2}{\chi_A} \phi_h^2$ (dashed black line) where $\chi_A$ only depends on $B$.  }
\label{fig:tauVSphih}
\end{figure}

Interestingly, the axial relaxation time $\tau_5$ displays a universal relation
\be
\label{universal}
(\tau_5 T)^{-1}\simeq\frac{2}{\chi_A}\phi_h^2\,,
\ee
where $\phi_h\equiv \phi(r_h)$ is the horizon value of the bulk scalar field. The coefficient $2/\chi_A$ depends only on the magnetic field $B$ and not on $\tilde{b}$, as demonstrated in the right panel of Fig.\ref{fig:tauVSphih}. This relation indicates that $\tau_5$, or equivalently $\Gamma$, depends only on a single IR quantity $\phi_h$. Nevertheless, it is important to notice that such an horizon value does not have a clear interpretation from the boundary field theory perspective, where it is a function of both $\tilde{M}$ and $\tilde{b}$. 
This universal relation is robust against variations of temperature $T$, magnetic field $B$, the chiral separation $b$, and the fermion mass $M$.  

Eq.\eqref{universal} can be obtained as follows.  
In holography, the longitudinal DC conductivity in the presence of a background magnetic field $B$ and axial relaxation reads 
\bea
\sigma_{\text{DC}}=\pi T+\frac{32B^2\alpha^2}{\pi^3 T^3 q^2\phi_h^2}\,,
\label{eq:DCholo}
\eea
where  
the second term is proportional to $B^2$ due to the axial charge relaxation and represents the phenomenon of negative magnetoresistance \cite{Landsteiner:2014vua, Jimenez-Alba:2015awa}. 
On the other hand, the longitudinal DC conductivity can also be calculated from chiral hydrodynamics \cite{Landsteiner:2014vua} at zero densities, 
\bea
\sigma_{\text{DC}}
%\sigma_V^{\text{tot}}
=\sigma_V%\pi T
+\frac{(8B\alpha)^2\tau_5}{\chi_A}\,,
\label{eq:DChydro}
\eea
where $\sigma_V\simeq \pi T$ since the above formulae are valid only when the magnetic field is weak and the axial relation time is long. 

Comparing the two equations, Eq.\eqref{eq:DCholo} and Eq.\eqref{eq:DChydro}, we conclude that
\bea
\Gamma=\tau_5^{-1}\simeq 2\pi^3T^3\frac{q^2\phi_h^2}{\chi_A}\,,
\label{eq:universal}
\eea
which is exactly the universal relation found numerically in Fig.\ref{fig:tauVSphih} where we work in the unit $\pi T=1$ and have set $q=1$. Note that this relation is valid only when $B/T^2\ll 1$ and $\tau_5 T\ll 1$. 

\begin{figure}[h]
\centering
\includegraphics[width=0.65\textwidth]{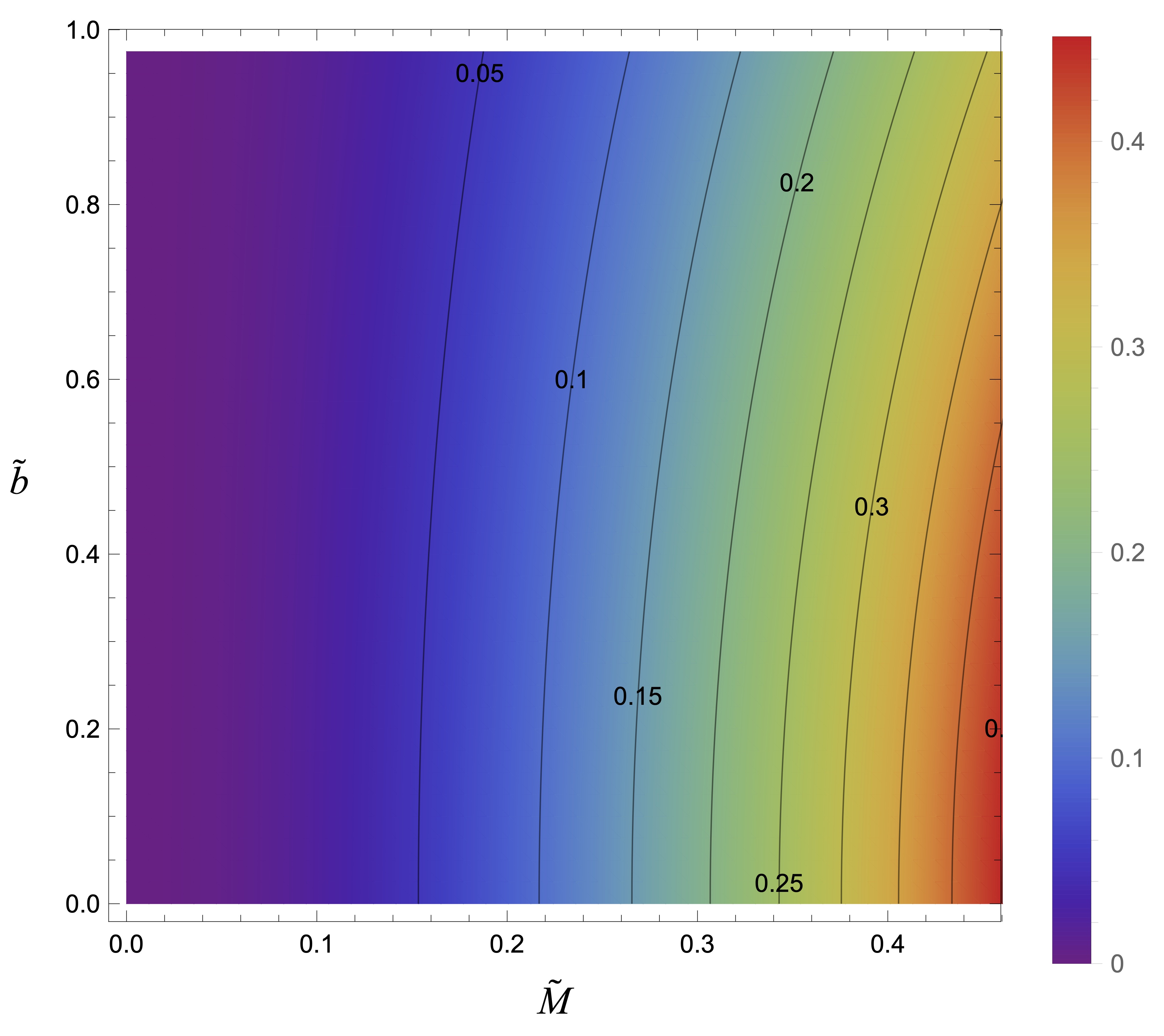}
\caption{Contour plot of $(\tau_5 T)^{-1}$ in the $\tilde{M}$-$\tilde{b}$ plane for $\tilde{B}=0.1$.  }
\label{fig:tauContour}
\end{figure}

To summarize our findings, we present a contour plot of $(\tau_5 T)^{-1}$ in the $\tilde{M}$-$\tilde{b}$ plane in Fig.\ref{fig:tauContour}. The changes in the rainbow color confirm that $(\tau_5 T)^{-1}$ increases with $\tilde{M}$, and the more and more dense contour lines indicate the quadratic increase. Finally, the vertical bending in the contour lines shows that chiral separation increases the lifetime of axial charge. 

%%%%%%%%%%%%%%%%%%%%%%%%%%%%%%%%%%%%%%%%%%%%%%%%%
\section{Coulomb interactions in the holographic Weyl semimetal model}
\label{Sec:screening}
%%%%%%%%%%%%%%%%%%%%%%%%%%%%%%%%%%%%%%%%%%%%%%%%%

Coulomb interactions are inevitably present in Dirac and Weyl semimetals, and impose corrections to the transport properties. As a concrete example, the susceptibility is defined from the second order variation of the free energy with respect to the charge density. Coulomb interactions between electrons and ions contribute to the free energy and hence the susceptibility gets screened.

In the presence of Coulomb interactions, the dynamical density-density correlation function in the random-phase approximation (RPA) is (see Appendix \ref{app:screening} for details)
\bea
\chi_{\text{sc}}(\omega, \vec{k})=\frac{\chi(\omega, \vec{k})}{1-\frac{\lambda_e}{|\vec{k}|^2}\chi(\omega, \vec{k})}\,,
\label{eq:scchi}
\eea
where the inverse momentum squared indicates the Coulomb potential in the momentum representation.  $\lambda_e$ is the electromagnetic coupling constant, which is left as a tunable parameter. 
On the left side, $\chi_{\text{sc}}(\omega, \vec{k})$ is the screened density-density correlation function in response to the external charge density $\rho_{\text{ext}}$. On the right side, $\chi(\omega, \vec{k})$ is the unscreened density-density correlation in response to the external gauge potential $v_{\text{ext}}$ that dictates the response of the system in absence of the Coulomb potential. 

Coulomb interactions have also important effects on the electric conductivity. In their presence, the (locally resolved) screened longitudinal conductivity $\sigma_{\text{sc}}$ is given in terms of the unscreened one $\sigma$ as
\bea
\sigma_{\text{sc}}(\omega,\vec{k})=\frac{\sigma(\omega, \vec{k})}{1-\frac{\lambda_e}{i\omega}\sigma(\omega, \vec{k})}\,,
\label{eq:scsigma}
\eea
which is also based on RPA. 

Coulomb interactions lead to different poles in the screened density-density and current-current correlation functions with respect to their unscreeed counterparts. The ratio between the unscreened susceptibility (conductivity) and the screened susceptibility (conductivity) defines the dynamical dielectric function, \textit{i.e.},
\bea
\epsilon(\omega,\vec{k})\equiv 1-\frac{\lambda_e}{|\vec{k}|^2}\chi(\omega, \vec{k})=1-\frac{\lambda_e}{i\omega}\sigma(\omega, \vec{k})\,,
\label{dielectric}
\eea
which plays a fundamental role in this discussion and which reduces to $1$ in absence of Coulomb interactions.
We can now anticipate the different meanings of poles of $\chi(\omega, \vec{k})$ and $\chi_{\text{sc}}(\omega, \vec{k})$ and their interpretation in the holographic side. To compute the poles of $\chi(\omega, \vec{k})$, we must take  Dirichlet boundary conditions for the vector gauge field at the boundary. This means that we switch off the external source of the charge density, \textit{e.g.}, $v_{\text{ext}}=0$. On the contrary, to compute the poles of $\chi_{\text{sc}}(\omega, \vec{k})$, we take the dynamical dielectric function to vanish
\bea
\epsilon(\omega,\vec{k})=0\,,
\label{dielectric}
\eea
which corresponds to the external charge density $\rho_{\text{ext}}$ to vanish, \textit{i.e.}, $\rho_{\text{ext}}=0$. 
In the zero wave-vector limit $\vec{k}\rightarrow 0$, $\epsilon(\omega=\Omega_{\text{pl}}, \vec{k}\rightarrow 0)=0$ gives the plasma frequency $\omega=\Omega_{\text{pl}}$. We refer to \cite{Gran:2017jht} for a nice discussion on this point in holography.
 
Here, we consider a dynamical gauge field in the boundary field theory by imposing mixed boundary conditions for the vector gauge field. We start from the total action for the gauge field  
\bea
\begin{split}
S_{\text{total}}^V&=\int d^5x\sqrt{-g}\Big[-\frac{1}{4}\mathcal{F}^2+\alpha\epsilon^{\mu\nu\rho\sigma\tau} A_\mu \mathcal{F}_{\nu\rho}\mathcal{F}_{\sigma\tau}\Big]\\
&+\int_{r=r_\infty} d^4x\sqrt{-\gamma}\Big[\frac{\text{log}~r}{4}\mathcal{F}^2-\frac{1}{4\lambda_e}\mathcal{F}^2+V_\mu J^\mu_{\text{ext}}\Big]\,,
\end{split}
\eea
where the last two terms are a boundary kinetic term and a Legendre transform. Then, $S^V_{\text{total}}$ leads to the equation of motion 
\bea
\Pi^\mu_V-\frac{1}{\lambda_e}\partial_\nu \mathcal{F}^{\mu\nu}+J^\mu_{\text{ext}}=0\,,
\label{eq:Maxwell}
\eea
where $\lambda_e$ is the electromagnetic coupling at the boundary since  we have fixed the bulk electromagnetic coupling to unit in the bulk. Considering the time component of Eq.\eqref{eq:Maxwell}, we find the electric permittivity (by comparing with Eq.\eqref{eq:Maxwelleqn})
\be 
\varepsilon_e=1/\lambda_e\,.
\ee 
The conjugate momentum $\Pi_V^\mu$ along the radial direction is given by
\bea
\Pi_V^\mu&=\displaystyle{\lim_{r\to\infty}}\frac{\delta S_{\text{ren}}}{\delta V_\mu}=\displaystyle{\lim_{r\to\infty}}\sqrt{-g}\left(\mathcal{F}^{\mu r}+4\alpha \epsilon^{r\mu\nu\rho\sigma}A_\nu\mathcal{F}_{\rho\sigma}\right)\,.
\label{Maxwell}
\eea
Here, we have taken the saddle point approximation and $S_{\text{ren}}$ is the renormalized on-shell action defined in Eq.\eqref{eq:onshell}. The radial conjugate momentum corresponds to the $U(1)$ vector current $\Pi_V^\mu=J_V^\mu$ with $\partial_\mu\Pi_V^\mu=0$ such that Eq.\eqref{Maxwell} is simply the Maxwell equation in the boundary field theory.

In terms of the boundary expansions ($r\rightarrow \infty$)
\bea
\begin{split}
v_t&=v_t^{(0)}+\frac{v_t^{(1)}}{r^2}\text{log}~r-\frac{v_t^{(2)}}{r^2}+...\,,\quad
v_z=v_z^{(0)}+\frac{v_z^{(1)}}{r^2}\text{log}~r+\frac{v_z^{(2)}}{r^2}+...\,,\\
a_t&=a_t^{(0)}+\frac{a_t^{(1)}}{r^2}\text{log}~r-\frac{a_t^{(2)}}{r^2}+...\,,\quad
a_z=a_z^{(0)}+\frac{a_z^{(1)}}{r^2}\text{log}~r+\frac{a_z^{(2)}}{r^2}+...\,,
\end{split}
\eea
we have
\bea
\begin{split}
\Pi_V^t&=2v_t^{(2)}+8\alpha B a^{(0)}_z\,,\quad
\partial_\mu\mathcal{F}^{\mu t}=k_z(\omega v_z^{(0)}+k_z v_t^{(0)})\,,\\
\Pi_V^z&=2v_z^{(2)}-8\alpha B a^{(0)}_t\,,\quad
\partial_\mu\mathcal{F}^{\mu z}=\omega(\omega v_z^{(0)}+k_z v_t^{(0)})\,.
\end{split}
\eea
The first-order variation of the conjugate momentum immediately follows as
\bea
\begin{split}
\delta\Pi_V^t&=2~\delta v_t^{(2)}+8\alpha B~\delta a^{(0)}_z\,,\quad
\delta \left(\frac{1}{\lambda_e}\partial_\mu\mathcal{F}^{\mu t}\right)=\frac{k_z}{\lambda_e}\left(\omega~\delta v_z^{(0)}+k_z~\delta v_t^{(0)}\right)\,,\\
\delta\Pi_V^z&=2~\delta v_z^{(2)}-8\alpha B~\delta a^{(0)}_t\,,\quad
\delta \left(\frac{1}{\lambda_e}\partial_\mu\mathcal{F}^{\mu z}\right)=\frac{\omega}{\lambda_e}\left(\omega~\delta v_z^{(0)}+k_z~\delta v_t^{(0)}\right)\,,
\end{split}
\eea
and therefore
\bea
\begin{split}
\delta J^t_{\text{ext}}&=-2~\delta v_t^{(2)}-8\alpha B~\delta a^{(0)}_z-\frac{k_z}{\lambda_e}\left(\omega~\delta v_z^{(0)}+k_z~\delta v_t^{(0)}\right)\,,\\
\delta J^z_{\text{ext}}&=-2~\delta v_z^{(2)}+8\alpha B~\delta a^{(0)}_t-\frac{\omega}{\lambda_e}\left(\omega~\delta v_z^{(0)}+k_z~\delta v_t^{(0)}\right)\,.
\end{split}
\eea
In terms of the U(1) gauge invariant quantity $v\equiv v_t+\frac{\omega}{k_z}v_z$, we obtain the variation rule
\bea
\begin{split}
\delta J^t_{\text{ext}}&=-\frac{2k_z^2}{\omega^2-k_z^2}\delta v^{(2)}-8\alpha B~\delta a^{(0)}_z-\frac{k_z^2}{\lambda_e}\delta v^{(0)}\,,\\
\delta J^z_{\text{ext}}&=-\frac{2\omega k_z}{\omega^2-k_z^2}\delta v^{(2)}+8\alpha B~\delta a^{(0)}_t-\frac{\omega k_z}{\lambda_e}\delta v^{(0)}\,.
\end{split}
\eea
Setting the external currents above to zero corresponds to a vanishing dynamical dielectric function, and gives us therefore the poles of the screened density-density correlation function.

%%%%%%%%%%%%%%%%%%%%%%%%%%%%%%%%%%%%%%%%%%%%%%%%%%%%
\section{Chiral magnetic waves (CMW) in the holographic Weyl semimetal model}
\label{sec:CMW}
%%%%%%%%%%%%%%%%%%%%%%%%%%%%%%%%%%%%%%%%%%%%%%%%%%

Let us first briefly review some general properties of the chiral magnetic waves. Since CMW appear both for finite densities $\mu\neq 0\,, \mu_5\neq0$ and zero densities $\mu =\mu_5=0$, for simplicity, we limit our analysis to the second simpler case.

In the absence of a magnetic field and axial charge relaxation, \textit{i.e.}, $B=\Gamma=0$, both the vector and axial charges diffuse at late time, corresponding to two independent hydrodynamic diffusive modes. Switching on a background magnetic field, $B\neq 0\,, \Gamma=0$, due to the axial anomaly, the two diffusive modes mix with each other via the CME and CSE and result in two propagating modes --- chiral magnetic waves \cite{Kharzeev:2010gd}. Instead, by switching off the magnetic field and considering the axial charge relaxation effect, \textit{i.e.} $B=0\,, \Gamma\neq 0$, the axial charge diffusive mode becomes a non-hydrodynamic gapped diffusive mode (described by a U(1) quasi-hydrodynamic theory \cite{Baggioli:2023tlc}) while the diffusive mode corresponding to vector charge fluctuations remains unchanged. 

In the presence of both the magnetic field and axial charge relaxation, \textit{i.e.} $B\neq0\,, \Gamma\neq 0$, the diffusive hydrodynamic mode arising from the conservation of the vector charge and the gapped non-hydro mode induced from the non-conserved axial charge display a complex interplay. This situation has been studied numerically in \cite{Jimenez-Alba:2014iia} but without considering the separation between the Weyl nodes which is relevant for the case of WSM.

\subsection{CMW in holographic WSM without Coulomb screening}
\label{CMW in WSM}

\begin{figure}[h]
\centering
\includegraphics[width=0.49\textwidth]{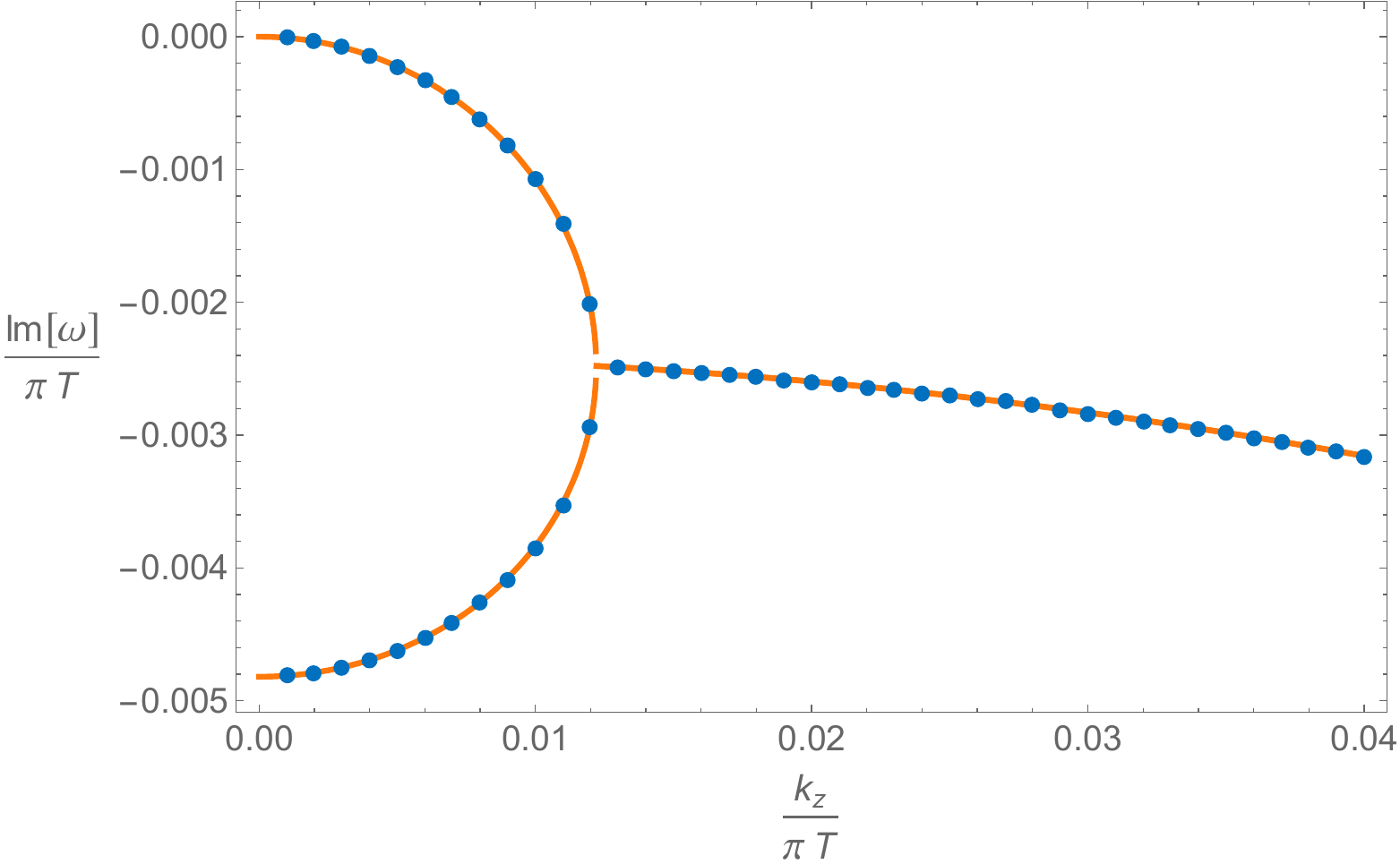}
\includegraphics[width=0.49\textwidth]{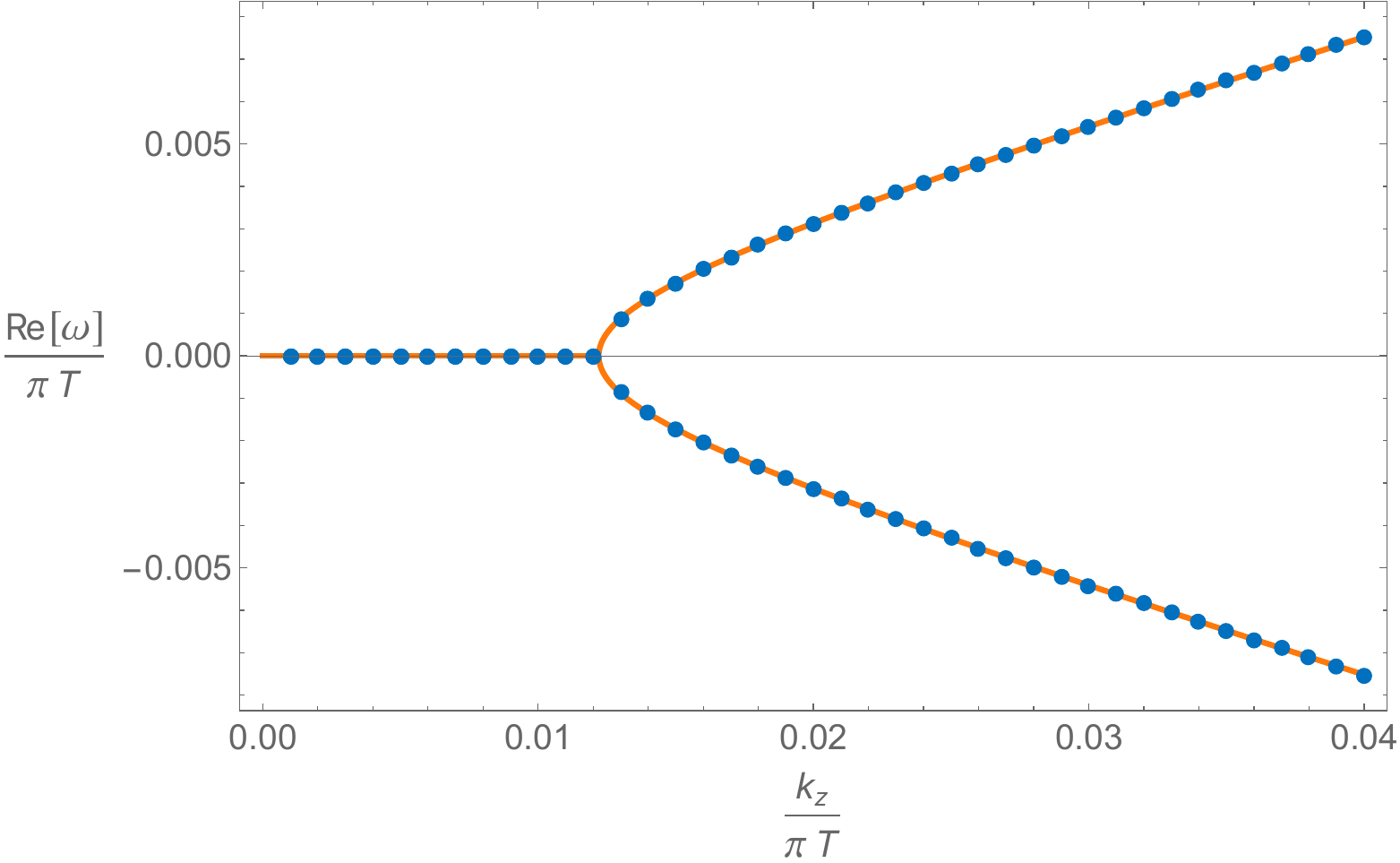}
\caption{The two lowest QNMs with $\tilde{B}=0.05\,, \tilde{b}=1\,, M/b=0.1$. The numerical results are shown by the numerical points while the prediction from quasi-hydrodynamics in Eq.\eqref{eq:CMWtype1} by the solid lines. }
\label{fig:CMWtype1}
\end{figure}

We first consider the simplest case without Coulomb screening. With the help of numerical techniques, we demonstrate that the dispersion relation of CMWs in holographic WSM is in perfect agreement with the predictions in Eq.\eqref{eq:CMWtype1} from chiral quasi-hydrodynamics. We provide an example in Fig.\ref{fig:CMWtype1}, where the orange solid lines represent the analytical predictions from Eq.\eqref{eq:CMWtype1}, while the blue dots are the numerical results from the holographic computations. We have checked that as long as $\tau_5 T\gg1$, Eq.\eqref{eq:CMWtype1} fits well the numerical results. 

The spectrum displays a hydrodynamic diffusive mode and a non-hydrodynamic transient mode that collide on the imaginary axes at a real value of the wave-vector $k_c$. Above such a value a real part of the dispersion develops and for $k\gg k_c$ a propagating wave re-emerges. Following \eqref{eq:CMWtype1}, the location of the critical point is given by
\bea
k_c=\frac{\Gamma \chi}{16\alpha B}\,,\quad\quad
\omega_c=-\frac{i\Gamma}{2}-iD\left(\frac{\Gamma\chi}{16\alpha B}\right)^2\,,
\eea
in the slow axial relaxation regime $\tau_5 T\gg1$. Here, $\chi$ and $D$ are only functions of $B$. When we fix $B, M$ and $T$, $\Gamma$ decreases monotonically with respect to chiral separation parameter $b$, as shown in Fig.\ref{fig:tauVSb}. As a consequence, chiral separation decreases the value of the critical wave-vector $k_c$ pushing the propagating regime towards smaller wave-vectors.

\subsection{CMW in holographic WSM with Coulomb screening}
\label{sec:CMWscreening}

Next, we take Coulomb interactions into consideration as a more realistic scenario for conducting semimetals. We recall the corresponding dispersion relation for the lowest excitations 
\bea
\omega_{\pm}=-\frac{i\Gamma}{2}-\frac{i\sigma}{2\varepsilon_e}-iDk_z^2\pm\frac{\sqrt{4\varepsilon_e\left(8\alpha B\right)^2\left(\chi+\varepsilon_e k_z^2\right)-(\sigma-\varepsilon_e\Gamma)^2\chi^2}}{2\varepsilon_e\chi}\,.
\label{eq:CMWtype3again}
\eea
One can notice that there are two timescales that might possibly destroy the hydrodynamic nature of the modes in Eq.\eqref{eq:CMWtype3again}: (I) $\tau_5^{-1}=\Gamma$ that comes from axial relaxation and (II) $\tau^{-1}=\sigma/\varepsilon_e$ that arises because of electric field relaxation due to Coulomb screening. Let us first assume $\tau_5\gg \tau$ so that axial relaxation can be ignored, \textit{i.e.} $\Gamma=0$. Then, we have
\bea
\omega_{\pm}=-\frac{i\sigma}{2\varepsilon_e}-iDk_z^2\pm\frac{\sqrt{4\varepsilon_e\left(8\alpha B\right)^2\left(\chi+\varepsilon_e k_z^2\right)-\sigma^2\chi^2}}{2\varepsilon_e\chi}\,. 
\label{eq:CMWtype2}
\eea
This equation is quite interesting. By tuning the magnetic field $B$ or the permittivity $\varepsilon_e$, the term in the square root can either be negative or positive as $k_z\rightarrow 0$, which is different from Eq.\eqref{eq:CMWtype1}, where that is always negative at $k_z=0$.

\begin{figure}[h]
\centering
\includegraphics[width=0.49\textwidth]{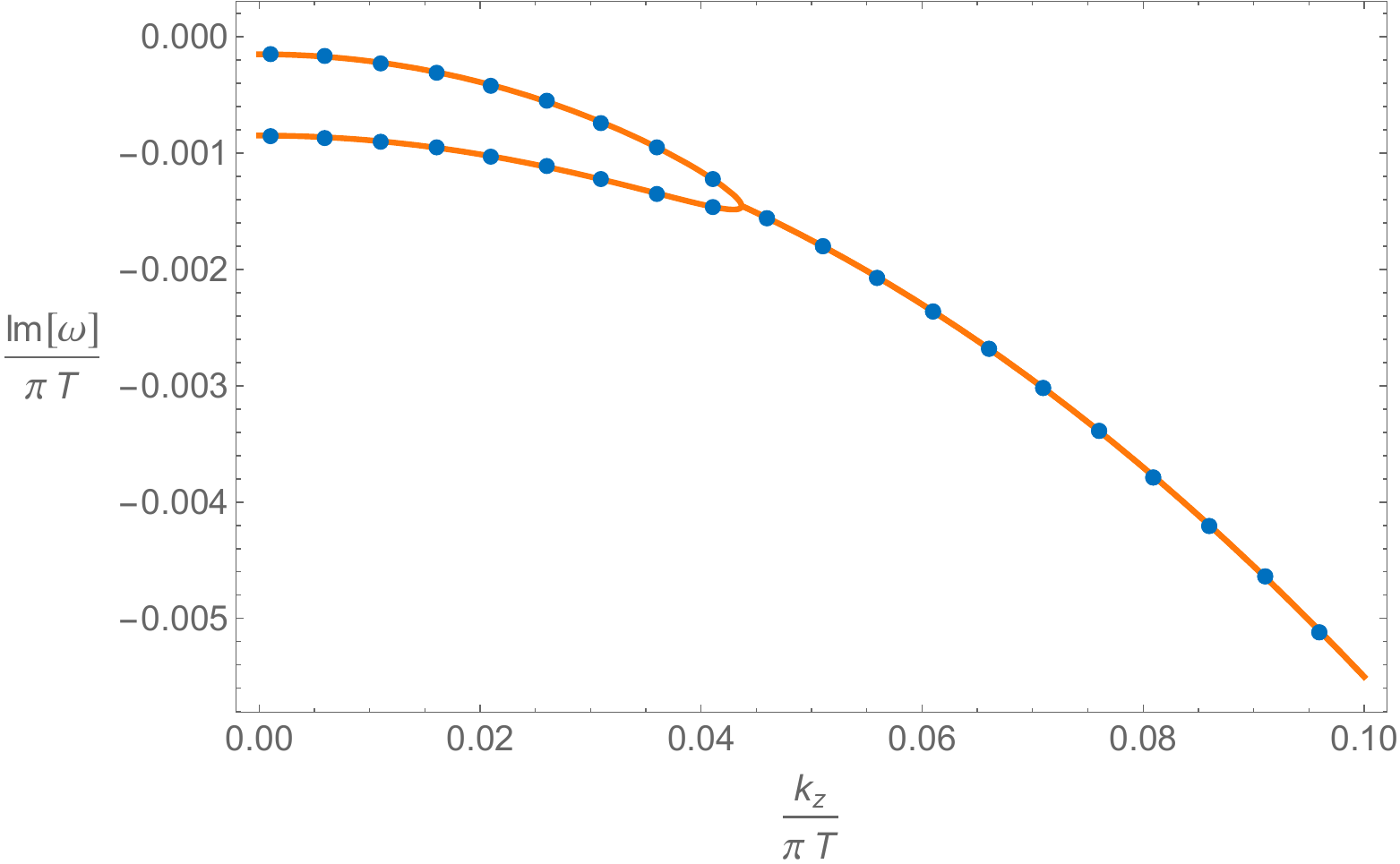}
\includegraphics[width=0.49\textwidth]{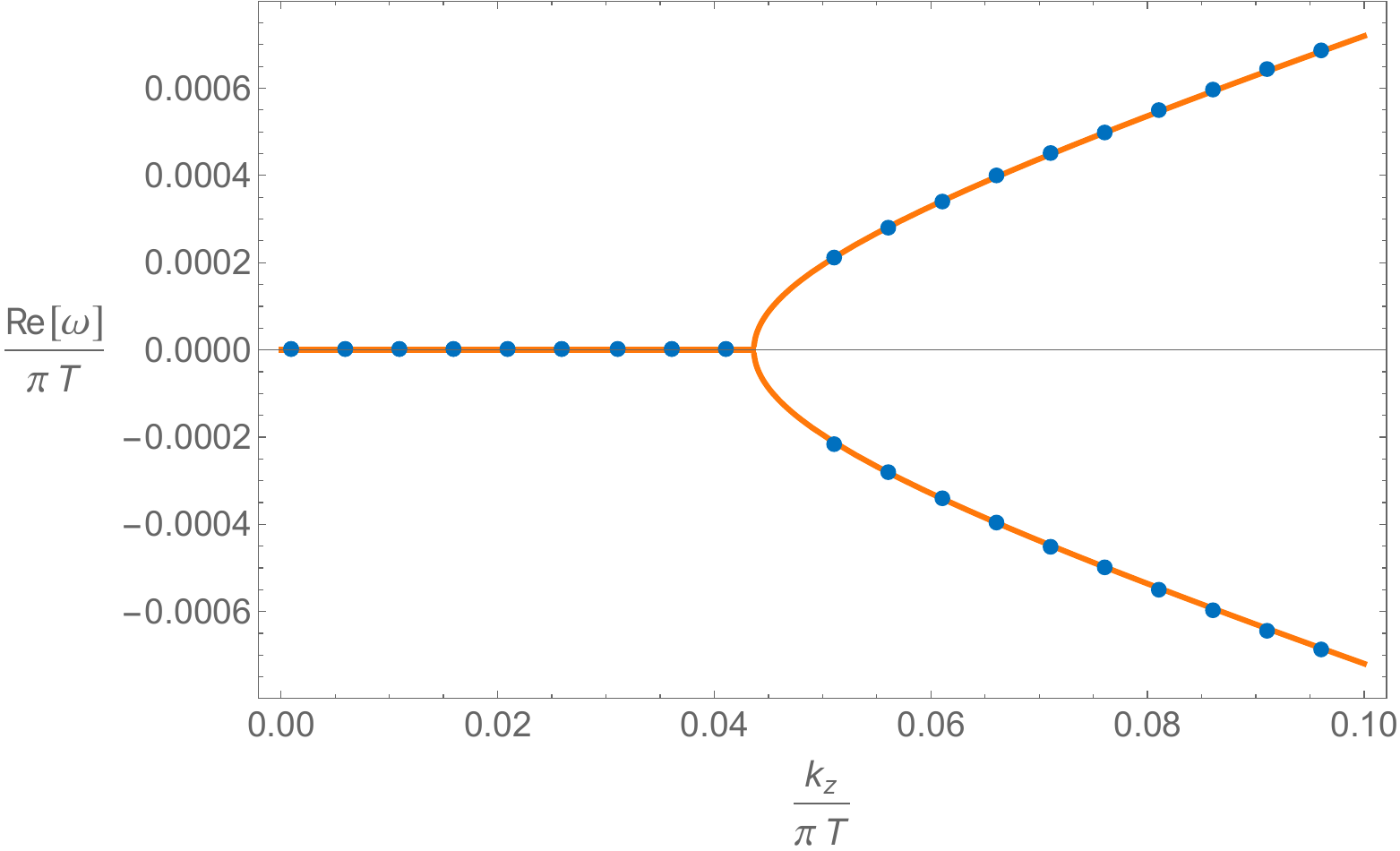}
\caption{Lowest modes with $\tilde{B}=0.002, \tilde{M}=\tilde{b}=0$ and $\lambda_e/(\pi T)=10^{-3}$. The numerical results are displayed by the numerical bullets while the quasi-hydrodynamical prediction from Eq.\eqref{eq:CMWtype2} by solid lines. }
\label{CMWtype2a}
\end{figure}

When the term inside the square root is negative, the two modes are purely imaginary in the small momentum regime, which is similar to Eq.\eqref{eq:CMWtype1}. However, now both of $\omega_{\pm}$ have an imaginary gap since even the vector charge dissipates at long wavelength due to screening effects. This structure of the modes is very similar to what found in the linear axion model for momentum relaxation \cite{Baggioli:2018nnp,Baggioli:2018vfc}. It has also been found and discussed in the context of holographic plasmons \cite{Gran:2018vdn,Baggioli:2019aqf,Baggioli:2019sio,Baggioli:2021ujk}, where nevertheless the dynamics are more complicated since they are a result of a three modes interaction. As an example, taking a sufficiently small $B$, Fig.\ref{CMWtype2a} demonstrates this qualitative difference in comparison with Fig.\ref{fig:CMWtype1}. Notice that in this regime the gap is still in wave-vector space and there is no finite real part below $k_c$.

\begin{figure}[h]
\centering
\includegraphics[width=0.49\textwidth]{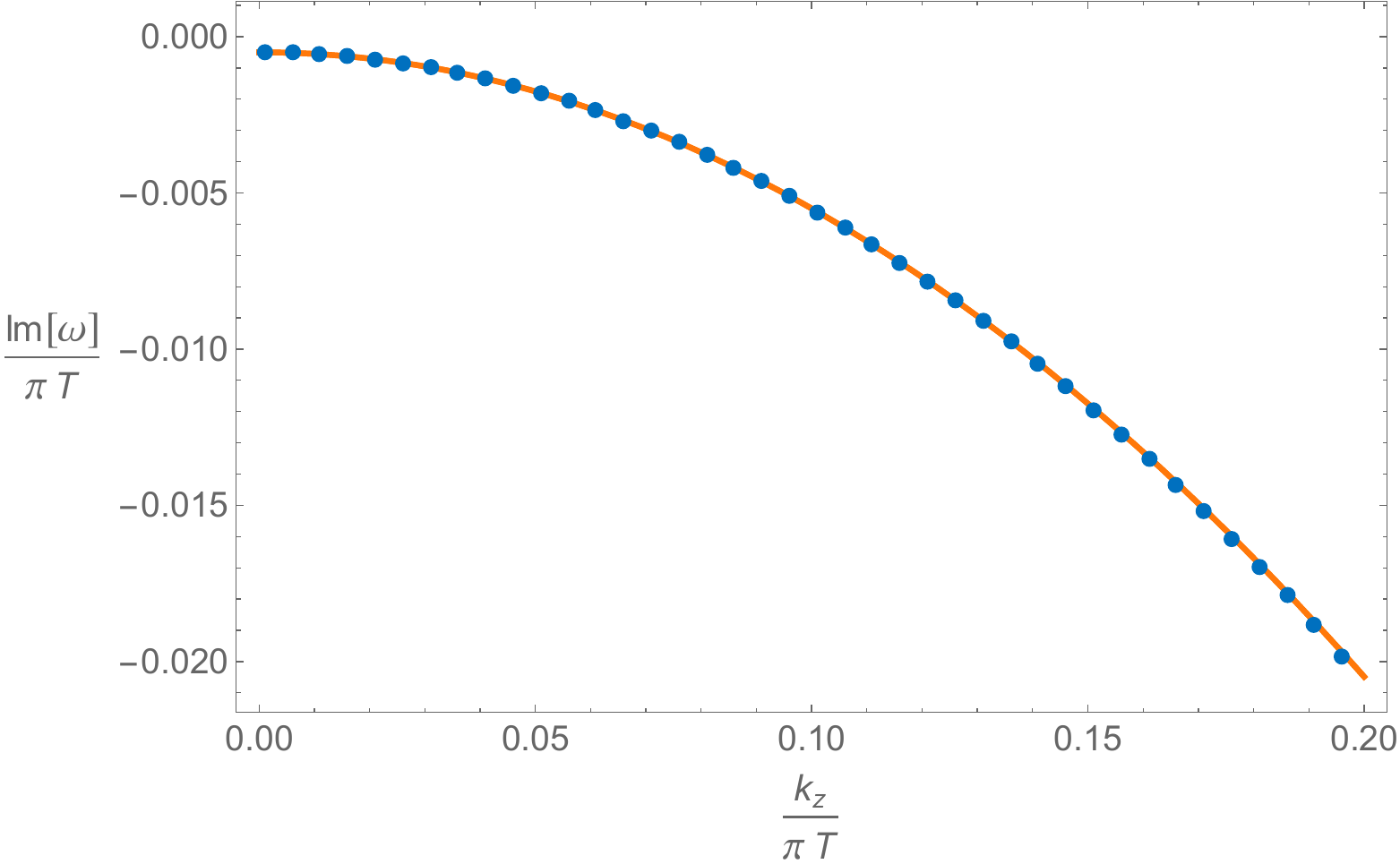}
\includegraphics[width=0.49\textwidth]{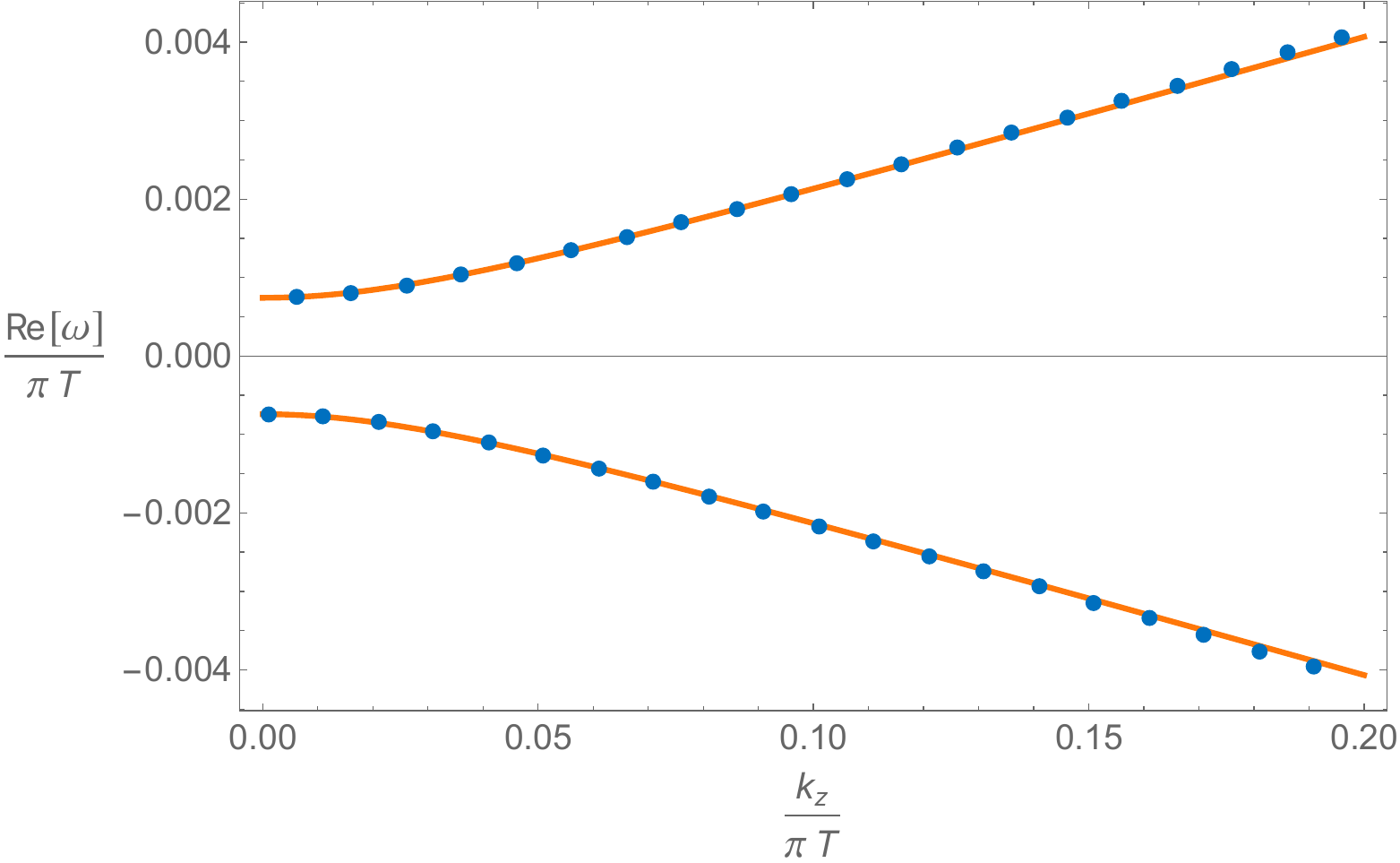}
\caption{Lowest modes with $\tilde{B}=0.005, \tilde{M}=\tilde{b}=0$ and $\lambda_e/(\pi T)=10^{-3}$. The numerical results are displayed by the numerical bullets while the quasi-hydrodynamical prediction from Eq.\eqref{eq:CMWtype2} by solid lines. }
\label{CMWtype2b}
\end{figure}

As the magnetic field $B$ increases and becomes larger than a critical value, the term in the square root in Eq.\eqref{eq:CMWtype2} can become positive for arbitrary values of the wave-vector $k$. As a result, the critical momentum $k_c$ does not exist anymore and the gap is transferred in frequency -- a mass gap. Once again, notice the similarities with \cite{Baggioli:2018vfc} (Fig.10 therein). This implies the emergence of a mass/frequency gap in as illustrated in Fig.\ref{CMWtype2b}. 

\begin{figure}[h]
\centering
\includegraphics[width=0.49\textwidth]{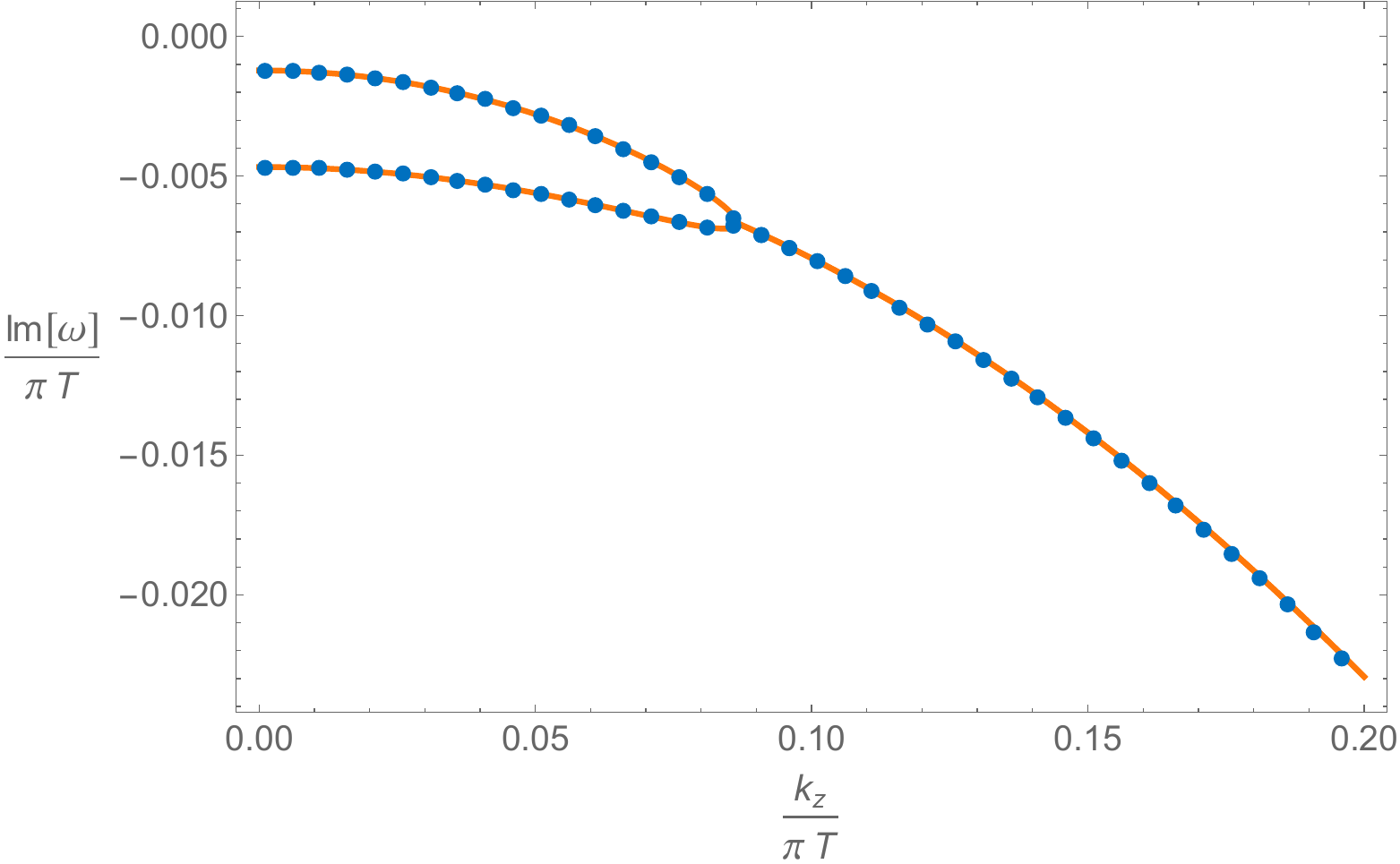}
\includegraphics[width=0.49\textwidth]{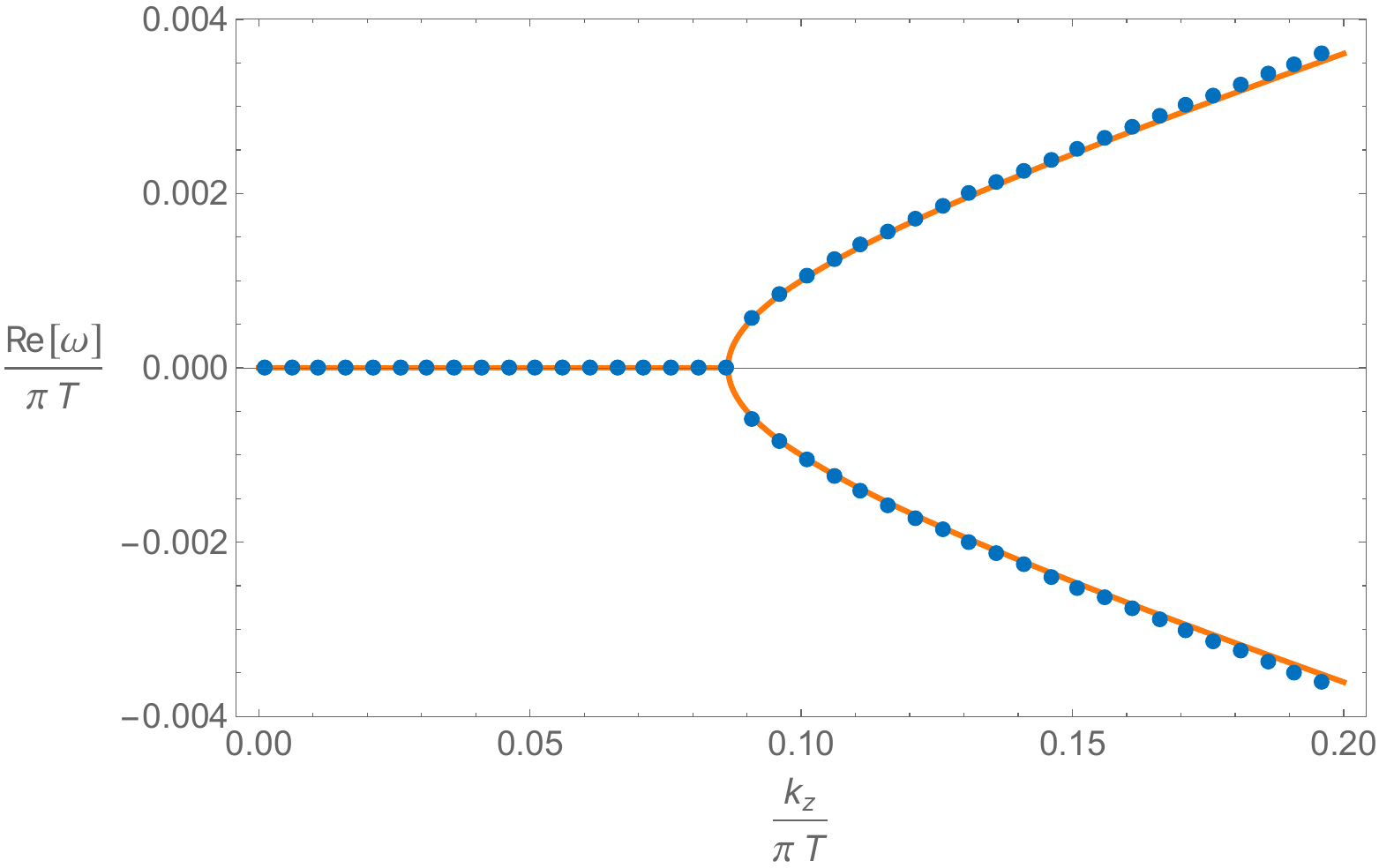}
\caption{Lowest QNMs with $\tilde{B}=0.005\,, \lambda_e/(\pi T)=10^{-3}\,, \tilde{b}=1$ and $M/b=0.1$.  The numerical results are represented by the numerical points while the prediction from quasi-hydrodynamics Eq.\eqref{eq:CMWtype3again} by solid lines. }
\label{CMWinWSM}
\end{figure}

Now, let us move beyond the limit in which axial relaxation is negligible and consider $\Gamma \neq 0$. As $\Gamma$ increases, the real parts in $\omega_{\pm}$ are suppressed. Reaching a critical $\Gamma$, the real parts vanish and the critical momentum $k_c$ further develops. We provide an example in Fig.\ref{CMWinWSM}, where the temperature $T$, magnetic field $B$, and permittivity $\varepsilon_e$ are the same as those in Fig.\ref{CMWtype2b}. The only difference is we switch on axial relaxation with $M/b=0.1$. One can find that axial relaxation forces the real part to disappear, and thus Fig.\ref{CMWinWSM} becomes qualitatively similar to Fig.\ref{CMWtype2a}, instead of Fig.\ref{CMWtype2b}. In other words, axial charge relaxation favours the scenario with a $k$-gap rather than that with a frequency gap.

\subsection{Properties of CMW and screened conductivity in holographic WSM}
\label{sec:realWSM}

\begin{figure}[H]
\centering
\includegraphics[width=0.43\textwidth]{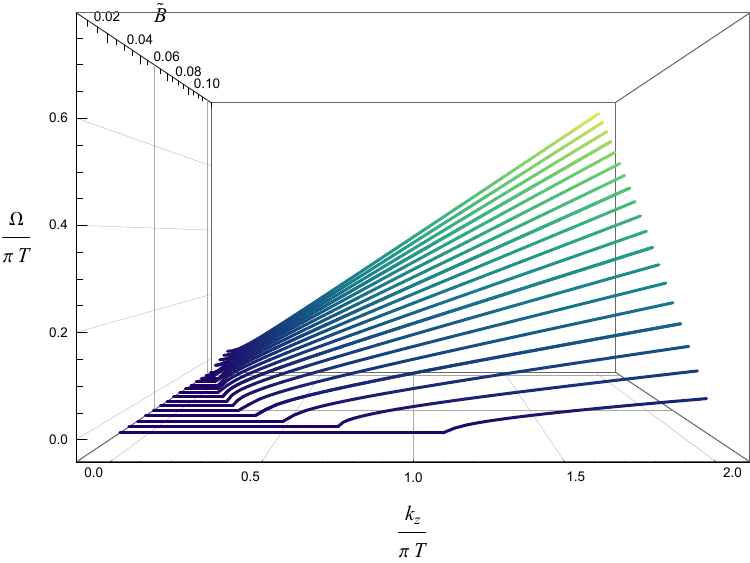}
\includegraphics[width=0.425\textwidth]{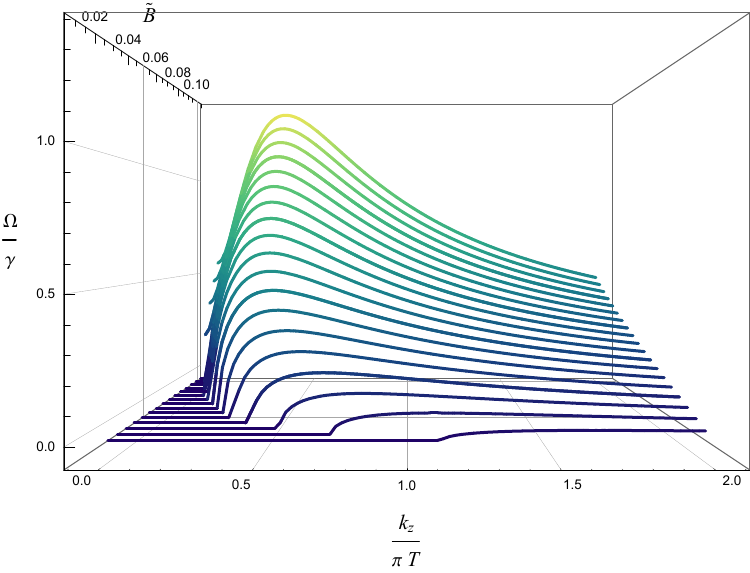}\\
\includegraphics[width=0.43\textwidth]{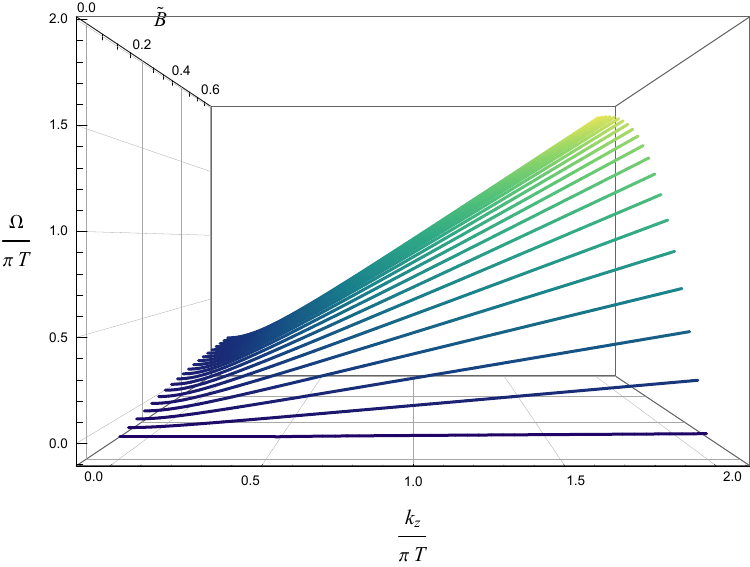}
\includegraphics[width=0.425\textwidth]{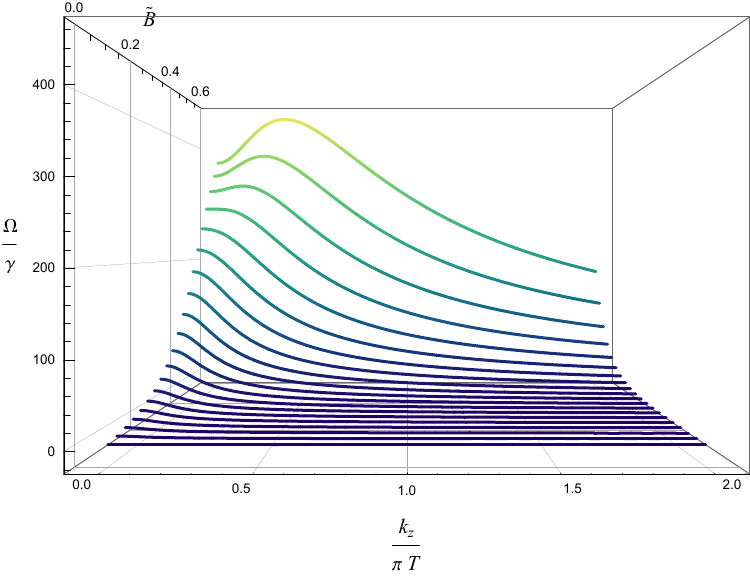}\\
\includegraphics[width=0.43\textwidth]{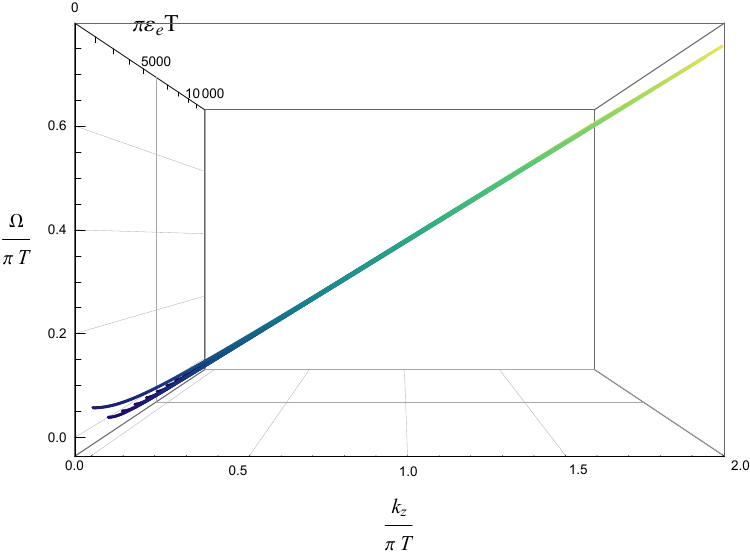}
\includegraphics[width=0.425\textwidth]{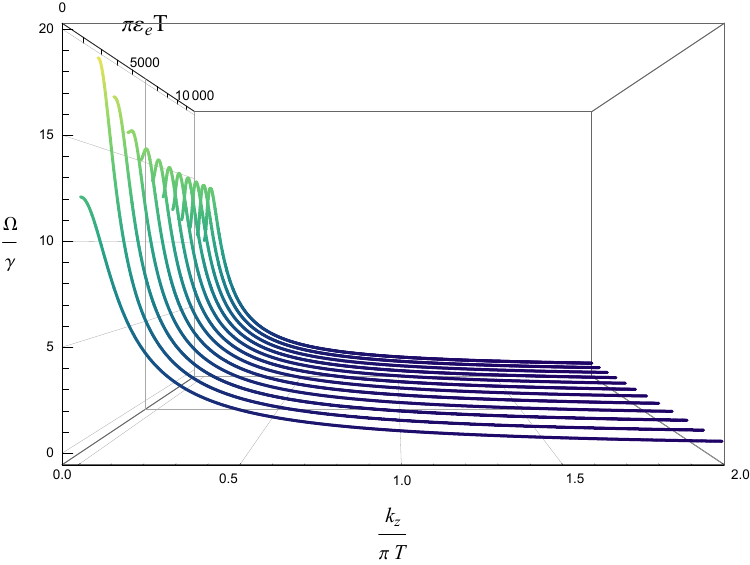}\\
\includegraphics[width=0.43\textwidth]{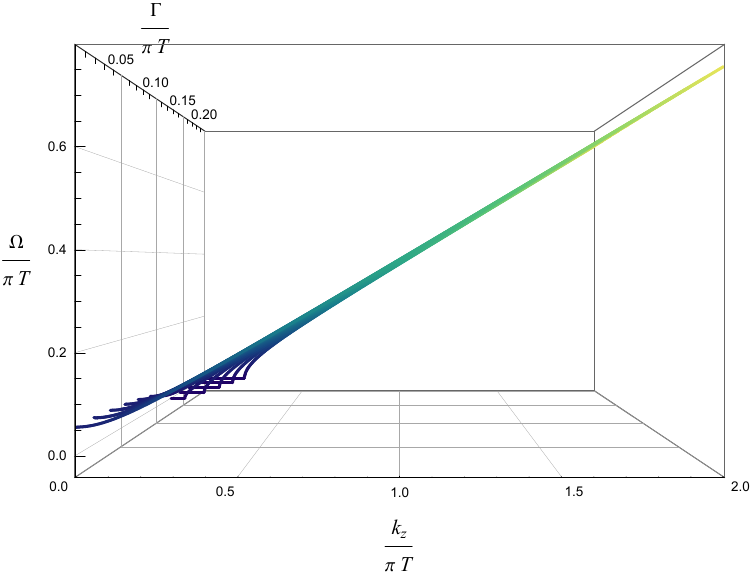}
\includegraphics[width=0.425\textwidth]{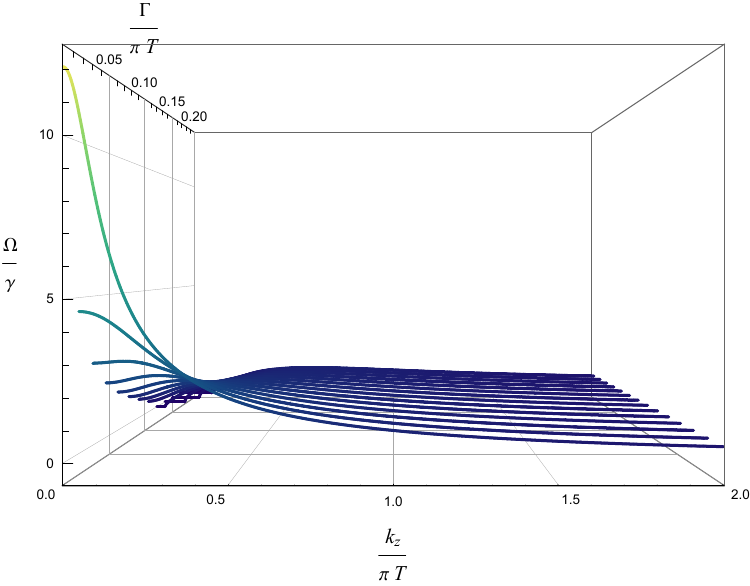}
\caption{Waterfall plots for the real part of the dispersion relation $\mathrm{Re}(\omega)\equiv -\Omega$ ({\bf Left}) and the ratio $\Omega/\gamma \equiv \mathrm{Re}(\omega)/\mathrm{Im}(\omega)$ ({\bf Right}) as a function of the wave-vector $k_z$ and the magnetic field (rows 1-2), permittivity (row 3), and axial relaxation (row 4) respectively. }
\label{fig:waterfall}
\end{figure}

We finally study the properties of the CMW in a strongly-coupled WSM. When the magnetic field is not too strong and the electric charge and axial charge slowly dissipate, the dispersion relations of the CMW are precisely described by the quasi-hydrodynamic theory presented in Section \ref{sec:hydro}. Therefore, in such limit, we can directly use the quasi-hydrodynamic expressions Eq.\eqref{eq:CMWtype3again} to illustrate our results in Fig.\ref{fig:waterfall}. 

As shown in Eq.\eqref{eq:CMWtype3again}, in general the dispersion of the lowest modes $\omega_\pm$ has a real part which corresponds to the characteristic energy of the excitation and a imaginary part describing its attenuation rate or lifetime. Without loss of generality, we focus on $\omega_-$ and define
\bea
\omega_-(k_z)=-\Omega(k_z)-i\gamma(k_z)\,,
\eea
where the minus sign in the real part indicates it propagates along the opposite direction with respect to the $z$ axis. $\Omega(k_z)$ and $\gamma(k_z)$ depend on the parameters $B, \varepsilon_e$ and $\Gamma$.
Importantly, we define the ratio $\Omega/\gamma$ as a good indicator to establish the nature of the excitations. For $\Omega/\gamma \gg 1$, the modes are underdamped and eventually propagating. On the contrary, for $\Omega/\gamma \ll 1$, the modes are overdamped and they are not able to propagate. This condition will obviously render their experimental detection very hard. 

The magnetic field $B$ plays the most crucial role in controlling the ratio $\Omega/\gamma$. In the small $B$ regime, there is a $k$-gap in the real part of the dispersion. As $B$ increases, the $k$-gap disappears and the absolute value of the real part increases with $B$. Correspondingly, $\Omega/\gamma$ at $k_z=0$ remains zero below a critical value of the magnetic field and can reach very high values when the magnetic field is strong, which indicates that the overdamped wave becomes underdamped.  As a concrete example, we take $\varepsilon_e=100$ and $\Gamma=1/1000$ (in unit $\pi T=1$) to show the effects of $B$ in the first and second rows in Fig.\ref{fig:waterfall}.

Both the energy of the CMW and the ratio $\Omega/\gamma$ are not very sensitive to the permittivity $\varepsilon_e$. As shown in the third row in Fig.\ref{fig:waterfall}, when we fix $B=0.1$ and $\Gamma=1/1000$ (in unit $\pi T=1$), the behavior of CMW remains qualitatively similar as $\varepsilon_e$ ranges from $10^2$ to $10^4$. 

The relaxation of the axial charge tends to destroy the CMW at small wave-vector, and consequently leads to a $k$-gap in the spectrum. For example, in the last row in Fig.\ref{fig:waterfall} we take $B=0.1$ and $\varepsilon_e=100$ (in unit $\pi T=1$) to demonstrate the effects of axial relaxation. It appears clear that a $k$-gap develops and then increases with $\Gamma$, while $\Omega/\gamma$ decreases monotonically to zero, which gradually makes the CMW overdamped and hence difficult to detect with experimental probes.  

In addition to the damped CMW, another observable that is accessible from experiments is the screened AC conductivity defined by
\bea
\sigma_{\text{sc}}(\omega,\vec{k}=0)=\frac{\sigma(\omega,\vec{k}=0)}{\epsilon(\omega,\vec{k}=0)}=\frac{\sigma(\omega,\vec{k}=0)}{1-\frac{\lambda_e}{i\omega}\sigma(\omega,\vec{k}=0)}\,,
\eea
where $\sigma(\omega,\vec{k}=0)$ is the bare AC conductivity computed from Eq.\eqref{eomAC} with infalling boundary conditions. Since the CMW is defined from the condition $\epsilon(\omega,\vec{k}=0)=0$, its signature should emerge in the conductivity when the imaginary part is small enough. As a result, $\sigma_{\text{sc}}(\omega,\vec{k}=0)$ is a good observable to detect the magnitude of $\Omega/\gamma$ and to observe the CMW when it is underdamped.

\begin{figure}[h]
\centering
\includegraphics[height=0.3\textwidth]{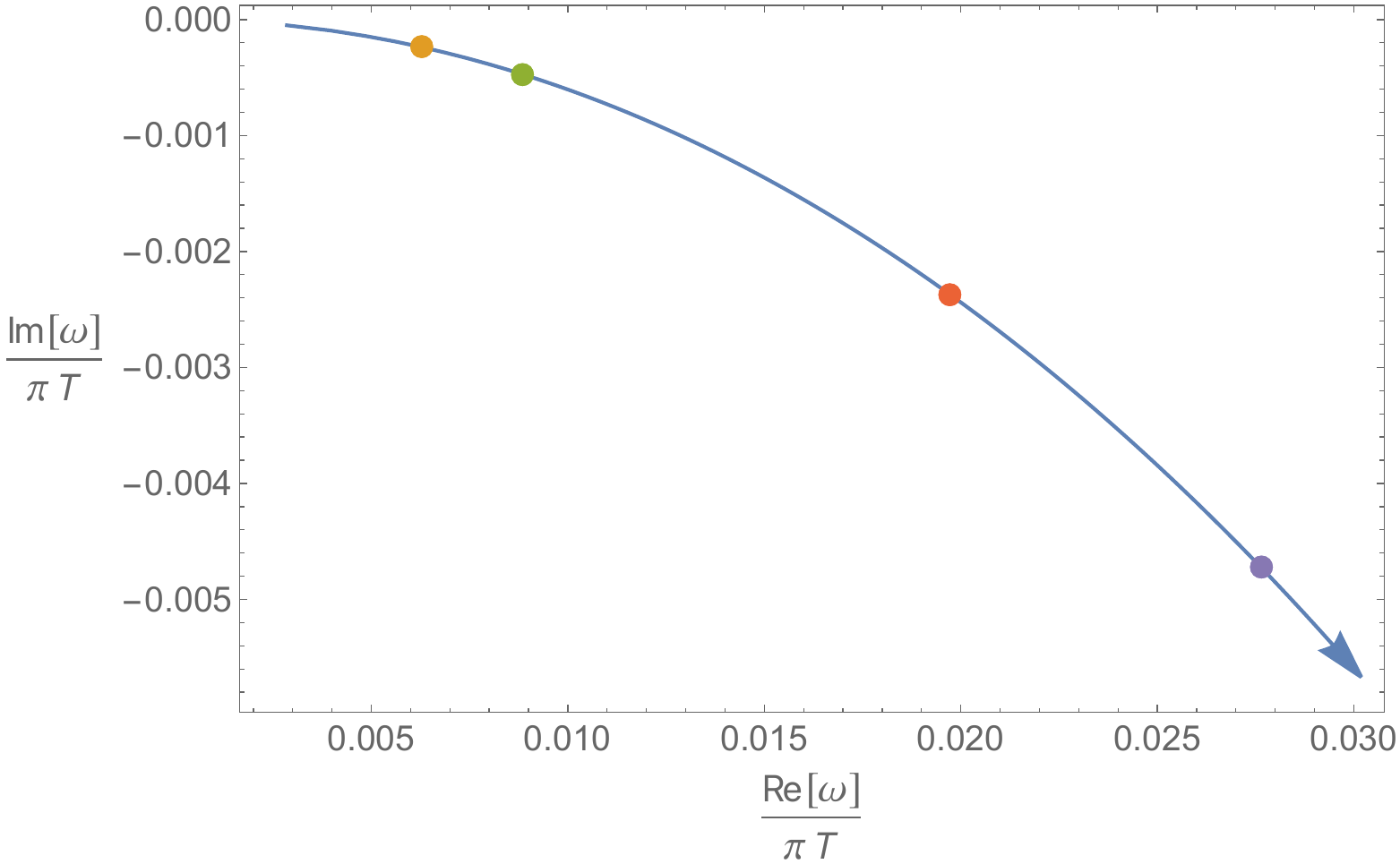}
\includegraphics[height=0.3\textwidth]{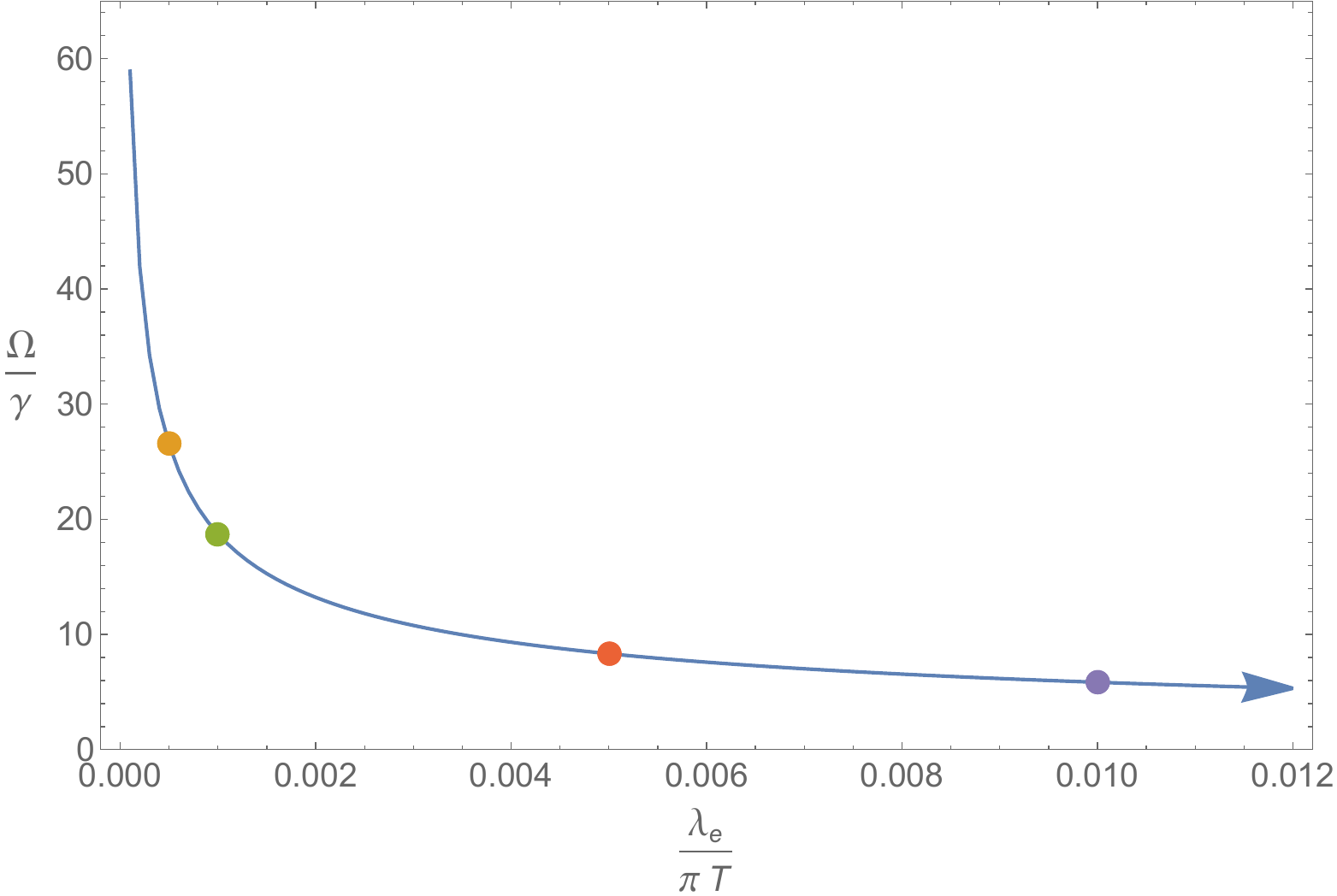}

\vspace{0.2cm}

\includegraphics[height=0.3\textwidth]{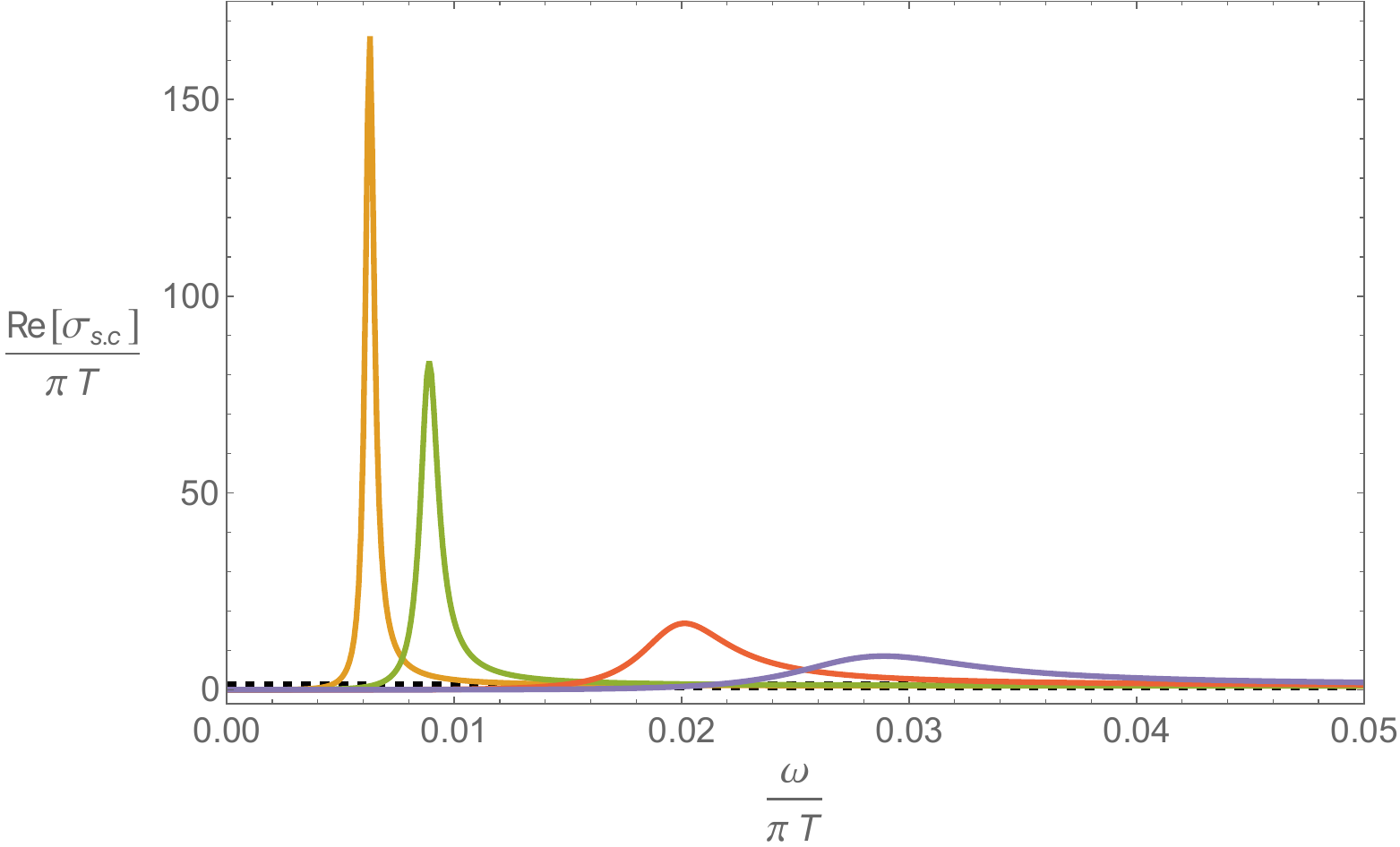}
\includegraphics[height=0.3\textwidth]{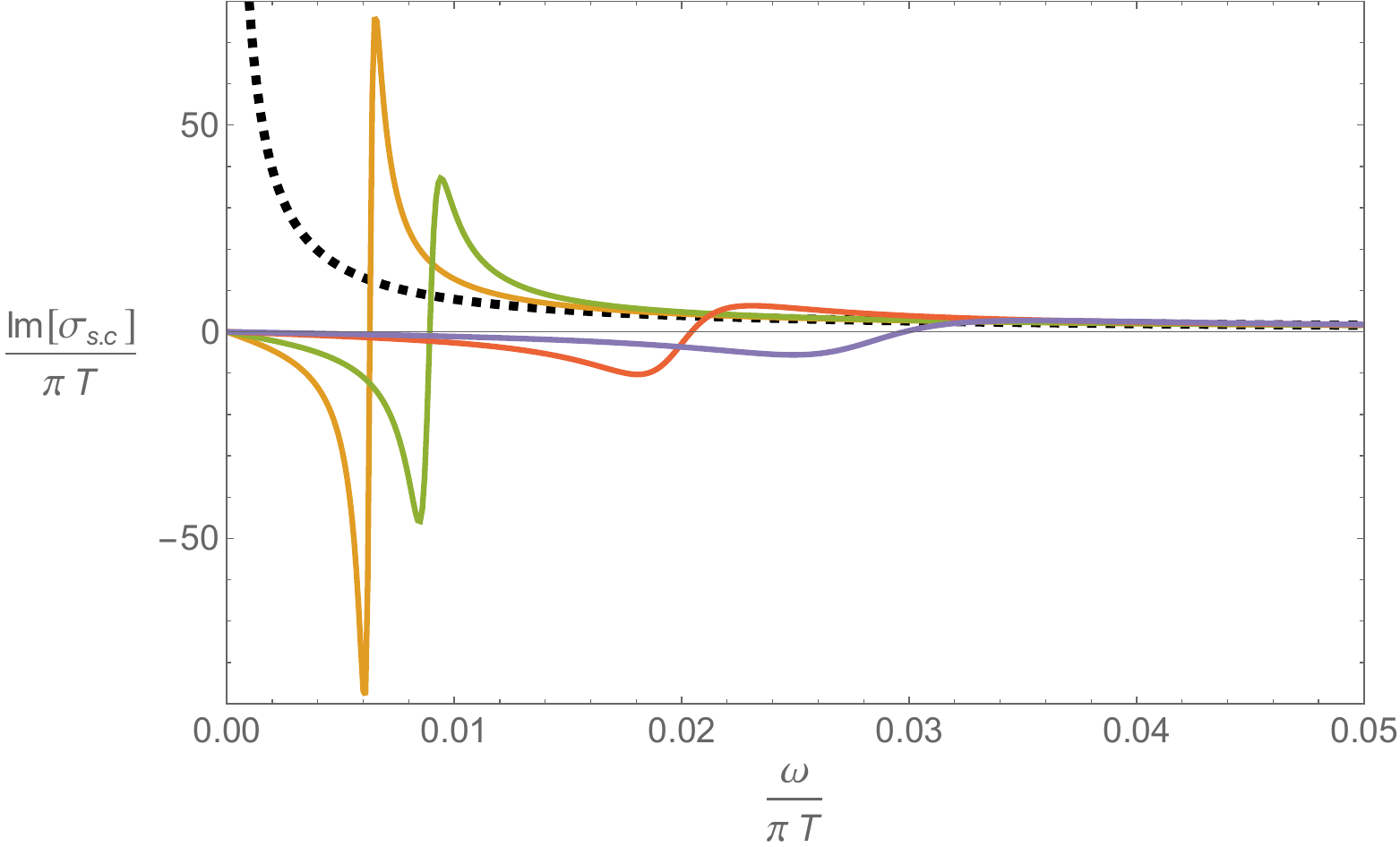}
\caption{{\bf Top:} Location shift of the lowest pole on the complex plane and the ratio $\Omega/\gamma$ by dialing the EM strength $\lambda_e$ in the direction of the arrow.  {\bf Bottom:} Real and imaginary parts of the optical screened conductivity $\sigma_{\text{sc}}$ for the corresponding parameters. }
\label{fig:AC}
\end{figure}

As shown in Fig.\ref{fig:AC}, both the real and imaginary parts of $\sigma_{\text{sc}}(\omega)$ present qualitatively different behaviors depending on whether $\Omega/\gamma$ is large or small. In the real Weyl semimetals, the axial charge can be nearly conserved \cite{burkov2015negative} and for simplicity, we take the $\Gamma=0$ limit in order to focus on the screening effect. The magnetic field is fixed as $\tilde{B}=0.05$, and we dial the strength of the Coulomb interactions, \textit{i.e.}, $\lambda_e=\varepsilon_e^{-1}$. Increasing $\lambda_e$, the location of the pole on the complex plane, defined from the condition $\epsilon(\omega,\vec{k}=0)=0$, moves towards larger values as shown by the blue curve in the upper left plot of Fig.\ref{fig:AC}. On the contrary, the ratio $\Omega/\gamma$ decreases with it as shown in the corresponding upper right plot. Regarding the screened conductivity, when the magnitude of $\Omega/\gamma$ is sufficiently large, the real part of $\sigma_{\text{sc}}$ presents a sharp resonance peak located at approximately, but not exactly, $\mathrm{Re}[\omega(0)]$. When the magnitude of $\Omega/\gamma$ diminishes, the width of the resonance peak increases and the response becomes incoherent and flat. A similar behavior is found in the imaginary part. Intuitively, the features found in the conductivity are just related to the nature of the corresponding excitation which needs to be underdamped to produce a nice peak feature in the conductivity, and therefore be detectable.

\begin{figure}[h]
\centering
\includegraphics[width=0.55\textwidth]{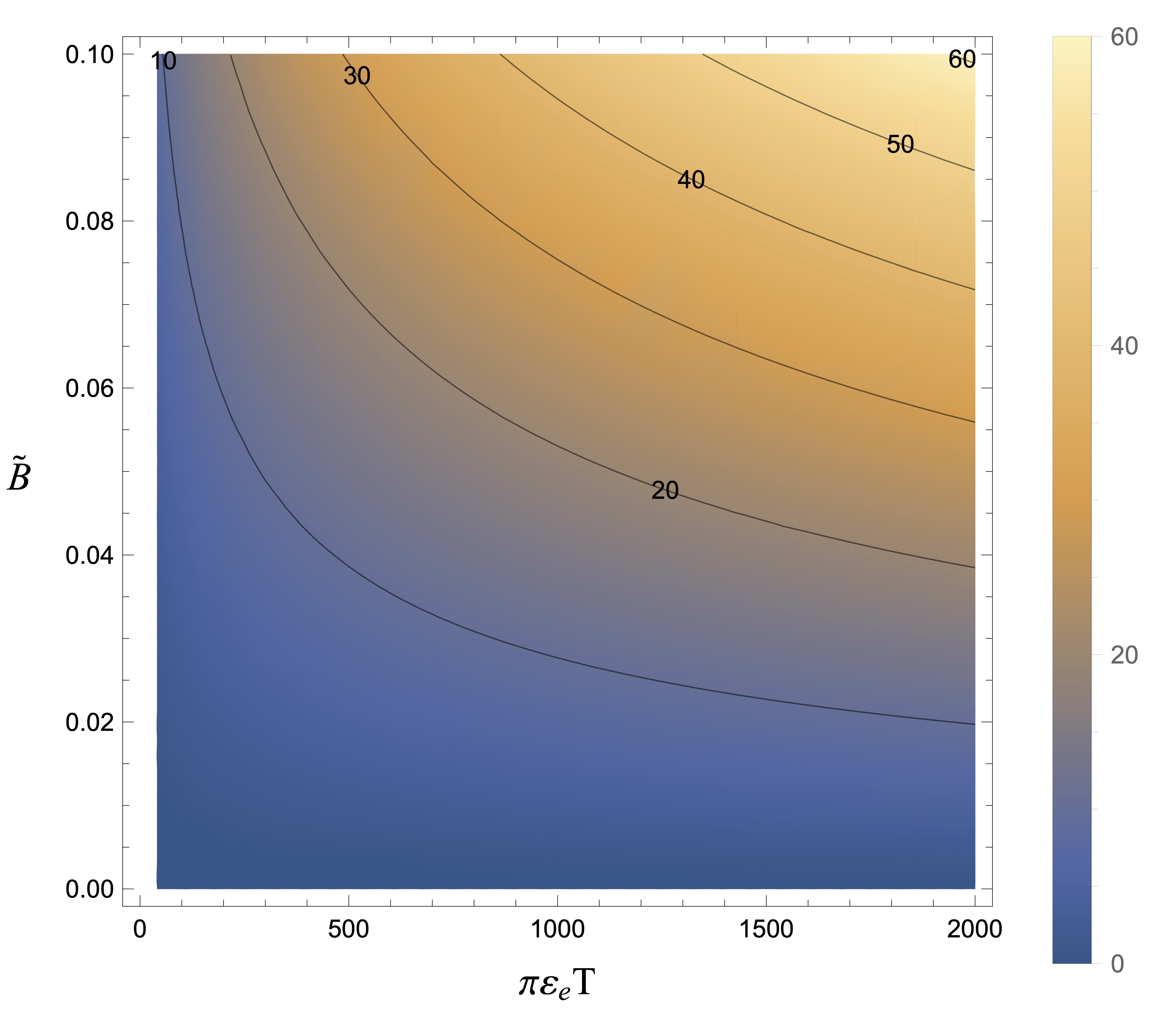}
\caption{Contour plot of $\Omega/\gamma$ at $\Gamma=0$ and $k_z=0$, in the $\varepsilon_e$-$\tilde{B}$ plane. }
\label{fig:E}
\end{figure}

We summarize this section with a contour plot of the ratio $\Omega/\gamma$ at $\Gamma=0$ and $k_z=0$, which is a reasonable approximation compared with the realistic Weyl semimetals. As shown in Fig.\ref{fig:E}, when the magnetic field is small, \textit{i.e.}, $\tilde{B}\lesssim 0.02$, the CMW is always overdamped with $\Omega/\gamma\lesssim 10$ and decays exponentially to zero at long wavelengths, $k_z\rightarrow 0$. In the strong Coulomb interaction limit $\lambda_e\rightarrow \infty$ ($\varepsilon_e\rightarrow 0$), the CMW is also overdamped due to charge decay. On the contrary, when the parameters are chosen to be above the contour line $\Omega/\gamma=20$, the CMW is underdamped and still propagates at long wavelength, leading to an observable peak in the real part of the screened AC conductivity $\sigma_\text{sc}$. 

\section{Conclusions}
\label{sec:conclusion}

Using hydrodynamics and bottom-up holography, we have performed a comprehensive study of the dispersion relation of collective chiral magnetic waves in strongly-coupled Weyl semimetals. In both frameworks, we have included the effects of axial charge relaxation and Coulomb interactions. Moreover, we have explicitly verified that the two methods give consistent results.

Using the holographic model as the underlying microscopic theory, we have first verified that the magnetic field does not affect the phase diagram in the probe limit. Then, we observed that the axial relaxation time $\tau_5$ becomes longer in the Weyl semimetal phase than in the trivial semimetal one, which indicates that the Weyl nodes separation tends to restore axial charge conservation. Consequently, we argued that a longer $\tau_5$ would facilitate the observation of chiral magnetic waves as underdamped collective excitations, making Weyl semimetals a possibly more promising experimental platform for their identification than heavy ion collisions. We also notice that in Weyl semimetal experiments, the behavior of negative longitudinal magnetoresistance \cite{burkov2015negative,zhang2016signatures} indicates the presence of a large axial relaxation time. 

Coulomb interactions, that are realized in holography using mixed boundary conditions for the bulk gauge field, modify the diffusion of vector charge fluctuations into a relaxation of the latter which is governed by the electric conductivity. However, Coulomb interactions also induce an energy gap (plasma frequency) that makes the dynamics of chiral magnetic waves more complicated. Now, different regimes can appear in which the dispersion relation of the chiral magnetic waves display a gap either in energy or in wave-vector, depending on which of the parameters dominate. We have confirmed that the hydrodynamic theory considered capture all these effects well and indeed is quantitatively consistent with the numerical data from holography. 

Finally, in order to verify whether chiral magnetic waves could be observable as underdamped excitations, we have investigated the ratio between the energy and relaxation rate, \textit{i.e.}, $\Omega/\gamma$, and focused on the effects of the magnetic field, Coulomb interactions, and axial relaxation that are the typical measurable parameters in Weyl semimetals. To provide a more direct test of the aforementioned ratio, we also computed the screened conductivity in the holographic Weyl semimetal model and showed a one-to-one correspondence with the damped propagating chiral magnetic wave. As expected, when the chiral magnetic wave is underdamped, we observe a peak in the screened conductivity $\sigma_{\text{sc}}$ that could be in principle directly accessible to experiments.  

Many questions are left to be explored.
It would be interesting to study the possibility of observing CMW in holographic nodal line semimetals \cite{Liu:2018bye, Liu:2020ymx, Rodgers:2021azg}, as well as to explore the observable effects of CMW in neutron stars \cite{Kaminski:2014jda,Hanai:2022yfh}, or even in our early universe \cite{Giovannini:1997eg,Kamada:2022nyt}. 

Moreover, it would be fruitful to investigate in more detail the dielectric response and the properties of the plasmon frequency with respect to magnetic field in topological semimetals \cite{xue2023electronic}, which might be another feasible method to detect CMW. Additionally, it is known that there might exist other types of waves in Weyl semimetals due to the torsional strain deformations, \textit{e.g.}, chiral sound wave \cite{Chernodub:2019lhw}, chiral density wave \cite{Baggioli:2020gpf}. It would be interesting to study these waves in the holographic Weyl semimetal model by introducing the effects of elasticity and strain (see \cite{RevModPhys.95.011001}).

\acknowledgments

We thank Karl Landsteiner and Igor Shovkovy for useful comments on a preliminary draft of this work. Y.A., M.B. and X.-M.W. acknowledge the support of the Shanghai Municipal Science and Technology Major Project (Grant No.2019SHZDZX01). M.B. acknowledges the sponsorship from the Yangyang Development Fund. Y.L. is supported by the National Natural Science Foundation of China grant No.11875083, 12375041. X.-M.W. is supported by the China Postdoctoral Science Foundation (Grant No. 2023M742296).

%\newpage
\appendix
\section{Equations of motion}
\label{App:A}
The action for the holographic WSM model is given by
\begin{align}
\begin{split}
  S=&\int d^5x \sqrt{-g}\bigg[\frac{1}{2\kappa^2}\Big(R+\frac{12}{L^2}\Big)\\
  &-\frac{1}{4e^2}\mathcal{F}^2-\frac{1}{4e^2}F^2  -(D_\mu\Phi)^*(D^\mu\Phi)-V(\Phi)  +\epsilon^{\mu\nu\rho\sigma\tau}A_\mu\bigg(\frac{\alpha}{3} \Big(F_{\nu\rho} F_{\sigma\tau}+3 \mathcal{F}_{\nu\rho}  \mathcal{F}_{\sigma\tau}\Big) \bigg)\bigg]\,,
  \end{split}
\end{align} 
where $V(\Phi)=m^2|\Phi|^2$ is the potential of the scalar. 
We consider the probe limit, \textit{i.e.}, the gravitational background is fixed to be the thermal AdS-Schwarzschild background. 
The equations of motion for the matter fields are 
\bea
\nabla_\nu \mathcal{F}^{\nu\mu}+2\alpha\epsilon^{\mu\tau\beta\rho\sigma} F_{\tau\beta}\mathcal{F}_{\rho\sigma}&=&0\,,\nonumber\\
\nabla_\nu F^{\nu\mu}+\epsilon^{\mu\tau\beta\rho\sigma} \Big[\alpha\big(F_{\tau\beta}F_{\rho\sigma}
+\mathcal{F}_{\tau\beta}\mathcal{F}_{\rho\sigma}\big)\Big]+i q\big[\Phi (D^\mu\Phi)^*-\Phi^*(D^\mu\Phi)\big]&=&0\,,\nonumber\\
D_\mu D^\mu\Phi-m^2\Phi&=&0\,.\nonumber
\eea 
In the manuscript, we have taken $2\kappa^2=1$ and $e^2=1$.

\subsection{Equations for the longitudinal fluctuations}
We consider the longitudinal sector and switch on the following fluctuations
\bea
\delta a_t\,,\quad \delta a_z\,,\quad \delta v_t\,,\quad \delta v_z\,,\quad \delta \phi_1\,,
\quad \delta \phi_2\,,
\eea
where $\delta \phi_1$ and $\delta\phi_2$ are the real and imaginary parts of the complex scalar fluctuation $\delta \Phi$, with the definition $\delta \Phi\equiv \delta \phi_1+i \delta\phi_2$.
We work in Fourier space by decomposing the fluctuations as
\bea
\begin{split}
\delta a_t&=a_t(r)e^{-i\omega t+ik_z z}\,,\quad
\delta a_z=a_z(r)e^{-i\omega t+ik_z z}\,,\\
\delta v_t&=v_t(r)e^{-i\omega t+ik_z z}\,,\quad
\delta v_z=v_z(r)e^{-i\omega t+ik_z z}\,,\\
\delta \phi_1&=\phi_1(r)e^{-i\omega t+ik_z z}\,,\quad
\delta \phi_2=\phi_2(r)e^{-i\omega t+ik_z z}\,.
\end{split}
\eea
In the probe limit, the second-order equations of motions for the fluctuations are
\bea
\begin{split}
a_t''+\frac{3a_t'}{r}-\left(\frac{k_z^2}{r^2u}+\frac{2q^2\phi^2}{u}\right)a_t+\frac{8\alpha Bv_z'}{r^3}-\frac{\omega k_z}{r^2u}a_z-\frac{2iq\omega\phi}{u}\phi_2&=0\,,\\
a_z''+\left(\frac{1}{r}+\frac{u'}{u}\right)a_z'+\left(\frac{\omega^2}{u^2}-\frac{2q^2\phi^2}{u}\right)a_z+\frac{8\alpha B}{ru}v_t'+\frac{\omega k_z}{u^2}a_t-\frac{4q^2A_z\phi}{u}\phi_1+\frac{2iqk_z\phi}{u}\phi_2&=0\,,\\
v_t''+\frac{3}{r}v_t'-\frac{k_z^2}{r^2u}v_t+\frac{8\alpha B}{r^3}a_z'-\frac{\omega k_z}{r^2u}v_z&=0\,,\\
v_z''+\left(\frac{1}{r}+\frac{u'}{u}\right)v_z'+\frac{\omega^2}{u^2}v_z+\frac{8\alpha B}{ru}a_t'+\frac{\omega k_z}{u^2}v_t&=0\,,\\
\phi_1''+\left(\frac{3}{r}+\frac{u'}{u}\right)\phi_1'+\left(\frac{\omega^2}{u^2}-\frac{m^2}{u}-\frac{k_z^2}{r^2u}-\frac{q^2A_z^2}{r^2u}\right)\phi_1-\frac{2qA_z}{r^2u}\left(q\phi a_z-ik_z\phi_2\right)&=0\,,\\
\phi_2''+\left(\frac{3}{r}+\frac{u'}{u}\right)\phi_2'+\left(\frac{\omega^2}{u^2}-\frac{m^2}{u}-\frac{k_z^2}{r^2u}-\frac{q^2A_z^2}{r^2u}\right)\phi_2-iq\phi\left(\frac{\omega a_t}{u^2}+\frac{k_za_z}{r^2u}\right)-\frac{2iqk_zA_z}{r^2u}\phi_1&=0\,.
\end{split}
\eea
There are also two first-order equations (constraints)
\bea
\begin{split}
\frac{\omega}{u}v_t'+\frac{k_z}{r^2}v_z'+\frac{8\alpha B}{r^3u}\left(k_za_t+\omega a_z\right)&=0\,,\\
\frac{\omega}{u}a_t'+\frac{k_z}{r^2}a_z'+\frac{8\alpha B}{r^3u}\left(k_zv_t+\omega v_z\right)+2iq\left(\phi\phi_2'-\phi'\phi_2\right)&=0\,.
\end{split}
\eea

\subsection{Equations for the transverse fluctuations}
\label{app:transverse}
In order to compute the anomalous Hall conductivity $\sigma_{\text{AHE}}$, we consider the transverse sector and switch on the fluctuations at $k_z=0$ as follows
\bea
\delta a_x\,,\quad \delta a_y\,,\quad
\delta v_x\,,\quad \delta v_y\,.
\eea
We consider only homogeneous perturbations, hence
\bea
\begin{split}
\delta a_x&=a_x(r)e^{-i\omega t}\,,\quad
\delta a_y=a_y(r)e^{-i\omega t}\,,\\
\delta v_x&=v_x(r)e^{-i\omega t}\,,\quad
\delta v_y=v_y(r)e^{-i\omega t}\,.
\end{split}
\eea
The equations for these perturbations read
\bea
\begin{split}
v_x''+\left(\frac{1}{r}+\frac{u'}{u}\right)v_x'+\frac{\omega^2}{u^2}v_x+\frac{8i\alpha\omega }{ru}A_z'v_y&=0\,,\\
v_y''+\left(\frac{1}{r}+\frac{u'}{u}\right)v_y'+\frac{\omega^2}{u^2}v_y-\frac{8i\alpha\omega }{ru}A_z'v_x&=0\,,\\
a_x''+\left(\frac{1}{r}+\frac{u'}{u}\right)a_x'+\left(\frac{\omega^2}{u^2}-\frac{2q^2\phi^2}{u}\right)a_x+\frac{8i\alpha\omega }{ru}A_z'a_y&=0\,,\\
a_y''+\left(\frac{1}{r}+\frac{u'}{u}\right)a_y'+\left(\frac{\omega^2}{u^2}-\frac{2q^2\phi^2}{u}\right)a_y-\frac{8i\alpha\omega }{ru}A_z'a_x&=0\,.\\
\end{split}
\eea
By changing variables to $v_{\pm}\equiv v_x\pm i\,v_y$, the equations simplify into
\bea
v_{\pm}''+\left(\frac{1}{r}+\frac{u'}{u}\right)v_{\pm}'+\frac{\omega^2}{u^2}v_{\pm}\pm\frac{8\alpha\omega }{ru}A_z'v_{\pm}&=0\,. 
\eea
In order to compute the DC $\sigma_{\text{AHE}}$, we expand $v_{\pm}$ around $\omega=0$ as 
\bea
v_{\pm}=\left(1-\frac{r_h^4}{r^4}\right)^{-\frac{i\omega}{4r_h}}\left(v_{\pm}^{(0)}+\omega v_{\pm}^{(1)}+...\right)\,.
\eea
After imposing the regularity condition near the horizon, we obtain the solutions $v_{\pm}^{(0)}=c_0$ and 
\bea
v_{\pm}^{(1)}=-\int_r^\infty dx\frac{c_0}{x^3 f}\left[\frac{ix^3f'}{4r_h}-ir_h\mp8\alpha\left(A_z-A_z(r_h)\right)\right]\,.
\eea
Therefore, the obtain the Green's function $G_{\pm}=\omega\left(\pm8\alpha(b-A_z(r_h))+ir_h\right)$ and 
\bea
\sigma_{xy}=\frac{G_+-G_-}{2\omega}=8\alpha\left(b-A_z(r_h)\right)\,,
\eea
which is the Hall conductivity for the covariant current. In order to consider the consistent current, we have to add the contribution form the Chern-Simons term and finally
\bea
\sigma_{\text{AHE}}=8\alpha b-\sigma_{xy}=8\alpha A_z(r_h)\,. 
\eea
In the presence of magnetic field $B$, $\sigma_{\text{AHE}}$ is a function of $B$ through the horizon value of $A_z$.

\subsection{On-shell action}
The on-shell action for the fluctuations is necessary to define the proper susceptibilities and conductivities. The  renormalized %on-shell 
action of the boundary field theory is given by 
\bea
S_\text{ren}=S+S_\text{bnd}
\label{eq:onshell}
\eea
with $S$ defined in Eq.\eqref{eq:holomodel} and 
\bea
\begin{split}
S _{\text{bnd}}&=\frac{1}{\kappa^2}\int_{r=r_\infty} d^4x \sqrt{-\gamma} K-\frac{1}{2\kappa^2}\int_{r=r_\infty} d^4 x\sqrt{-\gamma}\bigg[6+\frac{1}{2}R+...\bigg]
\\&~~~
+\frac{\log r}{4}\int_{r=r_\infty} d^4x \sqrt{-\gamma}\bigg[\mathcal{F}_{\mu\nu}\mathcal{F}^{\mu\nu}+F_{\mu\nu}F^{\mu\nu}+|D_m\Phi|^2+\frac{1}{3}|\Phi |^4\bigg]
\end{split}
\eea
where $\gamma_{ab}$ is the induced metric on the boundary. $K$ and $R$ are the extrinsic and intrinsic curvatures respectively. 

Substituting the fluctuations at $\omega\neq 0, k_z=0$ into the total renormalized action \eqref{eq:onshell}, we obtain the relevant part in the renormalized on-shell action
\bea
\begin{split}
S_{\text{o.s.}}\supset\int \frac{d\omega}{2\pi}dx^3\bigg[&-2a_t^{(0)}(-\omega)a_t^{(2)}(\omega)+2a_z^{(0)}(-\omega)a_z^{(2)}(\omega)\\
&-2v_t^{(0)}(-\omega)v_t^{(2)}(\omega)+2v_z^{(0)}(-\omega)v_z^{(2)}(\omega)\\
&+\mathcal{O}(\omega^2)+\text{contact terms}\bigg]\,.
\end{split}
\eea
Note that there is a factor $2$ which is crucial to properly define the susceptibilities and DC conductivities.

\section{Susceptibilities and conductivities}
\label{App:B}
The total axial charge susceptibility $\chi_A^{\text{tot}}$ is defined from the correlation function of the axial charge density, \textit{i.e.}, $\chi_A^{\text{tot}}=\langle\rho_A\rho_A\rangle_R|_{\omega=k=0}$.  
It can be obtained in holography by solving
\bea
a_t''+\frac{3}{r}a_t'-\left(\frac{64B^2\alpha^2}{r^4u}+\frac{2q^2\phi^2}{u}\right)a_t=0\,.
\eea
The asymptotic expansion  of $a_t$ near the boundary is given by
\bea
a_t(r)=a_t^{(0)}-\frac{q^2M^2a_t^{(0)}}{r^2}\text{ln}r-\frac{a_t^{(2)}}{r^2}+...\,,
\eea
hence $\chi_A^{\text{tot}}=\frac{2a_t^{(2)}}{a_t^{(0)}}$. %\textcolor{blue}{
$\chi_A$ in the constitutive equations is obtained by taking $\phi=0$ during the computation.%} 

The total vector charge susceptibility $\chi_V^{\text{tot}}$ is defined from the correlation function of the charge density, \textit{i.e.}, $\chi_V^{\text{tot}}=\langle\rho_V\rho_V\rangle_R|_{\omega=k=0}$, which can be extracted from
\bea
\begin{split}
v_t''+\frac{3}{r}v_t'+\frac{8B\alpha}{r^3}a_z'&=0\,,\\
a_z''+\left(\frac{1}{r}+\frac{u'}{u}\right)+\frac{8B\alpha}{ru}v_t'-\frac{2q^2\phi^2}{u}a_z&=0\,.
\end{split}
\eea
The UV expansions of the relevant bulk fields are given by
\bea
\begin{split}
a_z(r)&=a_z^{(0)}-\frac{q^2M^2a_z^{(0)}}{r^2}\text{ln}r+\frac{a_z^{(2)}}{r^2}+...\,,\\
v_t(r)&=v_t^{(0)}-\frac{v_t^{(2)}}{r^2}+...\,.
\end{split}
\eea
This implies that $\chi_V^{\text{tot}}=\frac{2v_t^{(2)}}{v_t^{(0)}}$ with the sourceless boundary condition $a_z^{(0)}=0$. $\chi_V$ in the constitutive equations is obtained by taking $\phi=0$ during the computation.

The total longitudinal DC axial conductivity $\sigma^A_{\text{DC}}$ is defined from the correlation function of the longitudinal axial current, \textit{i.e.}, $\sigma^A_{\text{DC}}=\text{Im}\frac{1}{\omega}\langle J_A^zJ_A^z\rangle_R|_{k_z=0,\omega\rightarrow 0}$, which can be obtained from
\bea
a_z''+\left(\frac{1}{r}+\frac{u'}{u}\right)+\left(\frac{\omega^2}{u^2}-\frac{2q^2\phi^2}{u}-\frac{64B^2\alpha^2}{r^4u}\right)a_z=0\,.
\eea
The UV expansion for $a_z$ reads
\bea
a_z(r)=a_z^{(0)}-\frac{q^2M^2a_z^{(0)}}{r^2}\text{ln}r+\frac{a_z^{(2)}}{r^2}+...\,,
\eea
and gives $\sigma^A_{\text{DC}}%\sigma_A^{\text{tot}}
=2\displaystyle{\lim_{\omega\to0}}\text{Im}\bigg[\frac{a_z^{(2)}}{a_z^{(0)}}\bigg]$. $\sigma_A$ in the constitutive equations is obtained by taking $\phi=0$ during the computation.

The total longitudinal DC electric conductivity $\sigma^V_{\text{DC}}$ can be computed analytically due to the existence of the radial conserved quantity when $\omega=0$. It can be solved from the coupled equations
\bea
\begin{split}
v_z''+\left(\frac{1}{r}+\frac{u'}{u}\right)+\frac{8B\alpha}{ru}a_t'&=0\,,\\
a_t''+\frac{3}{r}a_t'+\frac{8B\alpha}{r^3}v_z'-\frac{2q^2\phi^2}{u}a_t&=0\,.
\end{split}
\label{eq:eomV}
\eea
From the first equation in Eq.\eqref{eq:eomV} we have the radial conserved quantity $J=-ruv_z'-8B\alpha a_t$ satisfying $\partial_rJ=0$, which means $J|_{r\rightarrow\infty}=J|_{r\rightarrow r_0}\equiv j$ and $\sigma^V_{\text{DC}}$ can be computed at the horizon. Near the horizon, 
\bea
\begin{split}
v_z=-\frac{E}{4\pi T}\text{ln}(r-r_0)+...\,,\\
a_t=-\frac{2E(8B\alpha)}{(4\pi T)r_0^2(q\phi_h)^2}+...\,,
\end{split}
\label{eq:DChorizon}
\eea
and therefore 
\bea
\sigma^V_{\text{DC}}=\frac{j}{E}=\pi T+\frac{32B^2\alpha^2}{\pi^3 T^3 q^2\phi_h^2}\,.
\label{eq:11}
\eea
Alternatively, one can obtain the same result from numerices, by switching on a sub-leading term $a_1(r-r_0)$ in $a_t$ in Eq.\eqref{eq:DChorizon}, where $a_1$ is determined from the UV expansion
\bea
\begin{split}
a_t(r)&=a_t^{(0)}-\frac{q^2M^2a_t^{(0)}}{r^2}\text{ln}r-\frac{a_t^{(2)}}{r^2}+...\,,\\
v_z(r)&=v_z^{(0)}+\frac{v_z^{(2)}}{r^2}...\,,
\end{split}
\eea
with the sourceless boundary condition $a_t^{(0)}=0$. As a result, 
$\sigma^V_{\text{DC}}=\frac{2v_z^{(2)}}{E}$ is consistent with the analytic result Eq.\eqref{eq:11}.

\iffalse
The total longitudinal DC electric conductivity $\sigma^V_{\text{DC}}$ is defined from the correlation function of the longitudinal vector current, \textit{i.e.}, $%\sigma_V^{\text{tot}}
\sigma^V_{\text{DC}}
=\text{Im}\frac{1}{\omega}\langle J_V^zJ_V^z\rangle_{k_z=0,\omega\rightarrow 0}$, which can be obtained from the coupled equations
\bea
\begin{split}
v_z''+\left(\frac{1}{r}+\frac{u'}{u}\right)+\frac{8B\alpha}{ru}a_t'&=0\,,\\
a_t''+\frac{3}{r}a_t'+\frac{8B\alpha}{r^3}v_z'-\frac{2q^2\phi^2}{u}a_t&=0\,.
\end{split}
\eea
The UV expansion reads
\bea
\begin{split}
a_t(r)&=a_t^{(0)}-\frac{q^2M^2a_t^{(0)}}{r^2}\text{ln}r-\frac{a_t^{(2)}}{r^2}+...\,,\\
v_z(r)&=v_z^{(0)}+\frac{v_z^{(2)}}{r^2}...\,,
\end{split}
\eea
and therefore one has $%\sigma_V^{\text{tot}}
\sigma^V_{\text{DC}}
=2\displaystyle{\lim_{\omega\to0}}\text{Im}\bigg[\frac{v_z^{(2)}}{v_z^{(0)}}\bigg]$ with the sourceless boundary condition $a_t^{(0)}=0$. $\sigma_V$ in the constitutive equations is obtained by taking $\phi=0$ during the computation.
\fi

The axial charge relaxation effects on transport are negligible when $\tau T\rightarrow\infty$. In this limit, the equations can be solved analytically as a function of $B/T^2$, %\textcolor{blue}{
leading to the quantum critical conductivities in the presence of $B$, \textit{i.e.}%}
\bea
\sigma_A=\sigma_V=\sigma=\frac{\pi^2T}{8}\beta^2\text{sec}\left(\frac{\pi}{2}\sqrt{1-\beta^2}\right)\frac{\Gamma[\frac{3-\sqrt{1-\beta^2}}{4}]\Gamma[\frac{3+\sqrt{1-\beta^2}}{4}]}{\Gamma[\frac{5-\sqrt{1-\beta^2}}{4}]\Gamma[\frac{5+\sqrt{1-\beta^2}}{4}]}
\eea
with $\beta=\frac{8B\alpha}{\pi^2 T^2}$. 
When $\beta\ll 1$, we have $\sigma=\pi T (1-\frac{\log 2}{2}\beta^2+\mathcal{O}(\beta^4))$. 
%\MB{we dont define $\sigma_E$}

Finally, to compute the AC conductivity $\sigma_V(\omega)$ at zero momentum $\vec{k}=0$ without axial relaxation, \textit{i.e.}, $\phi=0$, we solve the ODE
\bea
v_z''+\left(\frac{1}{r}+\frac{u'}{u}\right)v_z'+\left(\frac{\omega^2}{u^2}-\frac{64\alpha^2B^2}{r^4u}\right)v_z=0\,
\label{eomAC}
\eea
with the infalling boundary condition of $v_z$ at the horizon.

\section{Coulomb screening, density-density correlation function and conductivities}
\label{app:screening}

In the following, we give a brief introduction to Coulomb screening effects on electric susceptibility (density-density correlation function) and electric conductivity (current-current correlation function). We follow the textbook \cite{girvin2019modern} and use SI units. 

For simplicity, we start from electrostatics in the jellium model. Switching on an external potential $v_{\text{ext}}(\vec{x})$, we can write down the energy functional $E[\rho]$ of the charge density $\rho(\vec{x})$ of the electrons with number density $n(\vec{x})$, and each electron carries charge $-e$ that defines the electric charge density $\rho(\vec{x})=-en(\vec{x})$,
\bea
\begin{split}
E[\rho]&=E_{\text{int}}[\rho]+\int d\vec{x}\,\rho(\vec{x})v_{\text{ext}}(\vec{x})\,,\\
E_{\text{int}}[\rho]&=E_{\text{int}}[\rho_0]+\frac{1}{2}\int d\vec{x}\int d\vec{x}'\delta \rho(\vec{x})\frac{\delta^2 E_{\text{int}}[\rho]}{\delta \rho(\vec{x})\,\delta \rho(\vec{x}')}\Big|_{\rho_0}\delta \rho(\vec{x}')\,+...
\end{split}
\eea
where in the Taylor expansion the linear term vanishes at equilibrium, $\rho(\vec{x})=\rho_0$. 

Minimizing the energy functional with respect to $\rho(\vec{x})$, we obtain 
\bea
\frac{\delta E[\rho]}{\delta \rho(\vec{x})}=\int d\vec{x}'\frac{\delta^2 E_{\text{int}}[\rho]}{\delta \rho(\vec{x})\,\delta \rho(\vec{x}')}\Bigg{|}_{\rho_0}\delta \rho(\vec{x}')+v_{\text{ext}}(\vec{x})=0
\eea
that defines the response function to the external source $v_{\text{ext}}$ 
\bea
\delta \rho(\vec{x})=\int d\vec{x}'\chi(\vec{x},\vec{x}')v_{\text{ext}}(\vec{x})\,,
\eea
and further the electric charge density-density correlation function as
\bea
\chi(\vec{x},\vec{x}')=-\Big[\frac{\delta^2 E_{\text{int}}[\rho]}{\delta \rho(\vec{x})\,\delta \rho(\vec{x}')}\Bigg{|}_{\rho_0}\Big]^{-1}\,.
\eea
Then, we perform the Fourier transformation and define the density-density correlation function in the momentum space
\bea
\chi(\vec{k})\equiv \int d\vec{x}\,\chi(\vec{x})e^{-i\vec{k}\cdot \vec{x}}\,, \quad\quad \text{with}\,\quad\quad \delta \rho(\vec{k})=\chi(\vec{k})v_{\text{ext}}(\vec{k})\,.
\eea

In order to distinguish the quantities with or without Coulomb screening, we use a normal symbol for the unscreened observables (non-interacting/bare quantities), while using a subscript ``sc'' for the screened ones (interacting/renormalized quantities).

We consider the energy functional
\bea
E_{\text{int}}[\rho]=\frac{1}{2}\int d\vec{x}\int d\vec{x}' \rho(\vec{x})\frac{\lambda_e}{4\pi|\vec{x}-\vec{x}'|}\rho(\vec{x}')+T_S[\rho]\,,
\eea
where the first term represents the energy due to the Coulomb potential with the strength tunable that is proportional to $\lambda_e$ and $T_S[\rho]$ the energy for non-interacting electrons. For the non-interacting electrons, the density-density correlation only comes from $T_S[\rho]$ that is defined as
\bea
\chi(\vec{x},\vec{x}')=-\Big[\frac{\delta^2 T_S[\rho]}{\delta \rho(\vec{x})\,\delta \rho(\vec{x}')}\Bigg{|}_{\rho_0}\Big]^{-1}\,.
\eea
Making a comparison between the density-density correlation function with and without Coulomb interactions in momentum space, we find
\bea
-\chi_{\text{sc}}^{-1}(\vec{k})=\frac{\lambda_e}{|\vec{k}|^2}-\chi^{-1}(\vec{k})\,.
\eea
After taking into account the frequency dependence to consider dynamical processes, we finally obtain the dynamical density-density correlation function in the random-phase approximation (RPA) that we ignore the exchange-correlation contribution in the energy functional and consider the effect of Coulomb interaction at Hartree level, 
%\MB{would be nice to clarify where the RPA approximation is taken and what means going beyond that}
\bea
\chi_{\text{sc}}(\omega, \vec{k})=\chi_{\text{RPA}}(\omega, \vec{k})=\frac{\chi(\omega, \vec{k})}{1-(\lambda_e/|\vec{k}|^2)\chi(\omega, \vec{k})}\,.
\label{eq:RPA}
\eea
On the left side, $\chi_{\text{sc}}(\omega, \vec{k})$ is the screened density-density correlation function in response to the external potential $v_{\text{ext}}$ with Coulomb potential. On the right side, $\chi(\omega, \vec{k})$ is the unscreened density-density correlation in response to the external potential $v_{\text{ext}}$ without Coulomb potential. 

We then discuss the different meanings of the poles of $\chi(\omega, \vec{k})$ and $\chi_{\text{sc}}(\omega, \vec{k})$ in holography, based on Eq.\eqref{eq:RPA}. To compute the poles of $\chi(\omega, \vec{k})$, we take the vanishing Dirichlet boundary condition for the field at the boundary. This means that we switch off the external source for the operator $\delta \rho$, e.g. $v_{\text{ext}}=0$. On the contrary, to compute the poles of $\chi_{\text{sc}}(\omega, \vec{k})$, we take the dynamical dielectric function to vanish
\bea
\epsilon(\omega,\vec{k})\equiv 1-(\lambda_e/|\vec{k}|^2)\chi(\omega, \vec{k})=0\,,
\label{dielectric}
\eea
which corresponds to the displacement field vanishing or the external charge $\rho_{\text{ext}}$ vanishing, \textit{i.e.} $\vec{D}=\rho_{\text{ext}}=0$. 
In the zero wave-vector limit $\vec{k}\rightarrow 0$, $\epsilon(\omega=\Omega_{\text{pl}}, \vec{k}\rightarrow 0)=0$ gives the plasma frequency $\omega=\Omega_{\text{pl}}$. In the finite wave-vector $\vec{k}$ regime, the dispersion relation of the plasma oscillations can be obtained from $\epsilon(\omega, \vec{k})=0$. 

Let us move to discuss the screening effects on the longitudinal conductivity. To start with, we recall the Maxwell's equations
\bea
\vec{\nabla}\cdot \vec{D}= \rho_{\text{ext}}\,,\quad\quad
\vec{\nabla}\cdot \vec{E}= (\rho_{\text{int}}+\rho_{\text{ext}}) \equiv  \rho_{\text{tot}}
\eea
where the electric displacement field $\vec{D}$ is related to the electric field $\vec{E}$ via the dielectric function as 
\bea
\vec{D}=\epsilon \vec{E}\,.
\eea
Here we should be careful to define the conductivity since we can express $\vec{J}\equiv\vec{J}_{\text{int}}$ with respect to $\vec{D}$ as well as $\vec{E}$ as
\bea
\vec{J}=\sigma_{\text{sc}} \vec{D}\,,\quad\quad
\vec{J}=\sigma \vec{E}\,.
\eea
In the first equation, the conductivity is defined with respect to the external field $\vec{D}$, then the Coulomb screening effect is included in the conductivity; while in the second equation, the current responds to the total electric field and the Coulomb screening effect is included in the electric field or the electric potential. Therefore, we can conclude that
\bea
\epsilon=\frac{\chi}{\chi_{\text{sc}}}=\frac{\sigma}{\sigma_{\text{sc}}}=\frac{\vec{D}}{\vec{E}}=\frac{V_{\text{ext}}}{V}\,,
\eea
where the electric potential $V_{\text{ext}}, V$ are defined from $\vec{D}=-\nabla V_{\text{ext}}$ and $\vec{E}=-\nabla V$ respectively. 

Next, we can relate the conductivity $\sigma$ with the dielectric function, via the continuity equation
\bea
\frac{\partial \rho}{\partial t}+\vec{\nabla}\cdot \vec{J}=0\,,
\eea
where both $\rho=-e n\equiv\rho_{\text{int}}$ and $\vec{J}\equiv\vec{J}_{\text{int}}$ are the responses inside the matter. 
After Fourier transformation, 
\bea
\label{eq:rho}
\rho(\omega,\vec{k})=\frac{1}{\omega}\vec{k}\cdot\vec{J}(\omega,\vec{k})
=\frac{1}{\omega}\sigma(\omega,\vec{k})(\vec{k}\cdot\vec{E}(\omega,\vec{k}))\,.
\eea
Recall that 
\bea
\label{eq:D}
\vec{D}(\omega,\vec{k})=\epsilon(\omega,\vec{k})\vec{E}(\omega,\vec{k})=\vec{E}(\omega,\vec{k})+\vec{P}(\omega,\vec{k})
\eea
where $\vec{P}$ is the polarization field satisfying $\vec{\nabla}\cdot \vec{P}=-\rho_{\text{int}}/\varepsilon_e$ with $\varepsilon_e=1/\lambda_e$. Combining Eq.\eqref{eq:rho} and Eq.\eqref{eq:D}, we finally obtain the relation
\bea
\epsilon(\omega, \vec{k})=1-\frac{\lambda_e}{i\omega}\sigma(\omega, \vec{k})
\eea
and therefore the relation between the screened longitudinal conductivity $\sigma_{\text{sc}}$ and the unscreened one $\sigma$ as
\bea
\sigma_{\text{sc}}(\omega,\vec{k})=\frac{\sigma(\omega, \vec{k})}{1-\frac{\lambda_e}{i\omega}\sigma(\omega, \vec{k})}\,,
\eea
which is also based on RPA. %\textcolor{red}{[$e^2$ or $\lambda_e$ is missing]}

\bibliographystyle{JHEP}
\bibliography{refs}

\end{document}